\documentclass[10pt,preprint]{aastex}
\usepackage{epsf}

\newcommand{\mpc}{\ensuremath{{\rm\,Mpc}}}

\newcommand{\hmpc}{\ensuremath{h^{-1}{\rm\,Mpc}}}
\newcommand{\ihmpc}{\ensuremath{h{\rm\,Mpc}^{-1}}}

\newcommand{\hgpcC}{\ensuremath{h^{-3}{\rm\,Gpc^3}}}

\newcommand{\msun}{\ensuremath{\rm M_\odot}}

\newcommand{\bfs}{\ensuremath{\vec{s}}}
\newcommand{\bfr}{\ensuremath{\vec{r}}}
\newcommand{\bfu}{\ensuremath{\vec{u}}}

\newcommand{\bfk}{\ensuremath{\vec{k}}}

\newcommand{\upair}{u_\parallel^{\rm pair}}

\newcommand{\beq}{\begin{equation}}
\newcommand{\eeq}{\end{equation}}
\newcommand{\beqa}{\begin{eqnarray}}
\newcommand{\eeqa}{\end{eqnarray}}

\newcommand{\tableskip}{\\[-6pt]}
\newcommand{\singleline}{\tableskip\hline\tableskip}
\newcommand{\doubleline}{\tableskip\hline\hline\tableskip}

\begin{document}

\shorttitle{Baryon Acoustic Oscillations}

\title{On the Robustness of the Acoustic Scale \\
in the Low-Redshift Clustering of Matter}
\author{
Daniel J.\ Eisenstein\altaffilmark{1,3}, 
Hee-Jong Seo\altaffilmark{1},
Martin White\altaffilmark{2} 
}

\begin{abstract}
We discuss the effects of non-linear structure formation on the 
signature of acoustic oscillations in the late-time galaxy distribution.
We argue that the dominant non-linear effect is the differential motion
of pairs of tracers separated by 150 Mpc.  These motions are driven
by bulk flows and cluster formation and are much smaller than the
acoustic scale itself.
We present a model for the non-linear evolution based on the 
distribution of pairwise Lagrangian displacements that provides a quantitative
model for the degradation of the acoustic signature, even for
biased tracers in redshift space.  The Lagrangian displacement 
distribution can be calibrated with a significantly smaller set
of simulations than would be needed to construct a precise power
spectrum.
By connecting the acoustic signature in the Fourier basis with that in the
configuration basis, we show that the acoustic signature is more robust
than the usual Fourier-space intuition would suggest because the beat
frequency between the peaks and troughs of the acoustic oscillations
is a very small wavenumber that is well inside the linear regime.
We argue that any possible shift of the acoustic scale is related
to infall on 150 Mpc scale, which is $O(0.5\%)$ fractionally at 
first-order even at $z=0$.  For the matter, there is a first-order 
cancellation such that the mean shift is $O(10^{-4})$.  However,
galaxy bias can circumvent this cancellation and produce a 
sub-percent systematic bias.
\end{abstract}


\altaffiltext{1}{Steward Observatory, University of Arizona,
		933 N. Cherry Ave., Tucson, AZ 85121}
\altaffiltext{2}{Departments of Physics and Astronomy,
                 University of California, Berkeley, CA 94720}
\altaffiltext{3}{Alfred P.~Sloan Fellow}


\section{Introduction}

The imprint in the late-time clustering of matter from the baryon acoustic
oscillations in the early Universe \citep{PeeYu70,Sun70,DZS78} has emerged as
an enticing way to measure the distance scale and expansion history of the
Universe 
\citep{EHT98,CHHJ01,Eis03,Bla03,Hu03,Seo03,Lin03,Mat04,Ame04,Bla05,Gla05,Dol06}. 
The distance that acoustic waves can propagate in the first million years of
the Universe becomes a characteristic scale, measurable not only in the CMB
anisotropies \citep{Mil99,deB00,Han00,Hal01,Net02,Ben03} but also in the late-time
clustering of galaxies \citep{Col05,Eis05}.
This scale can be computed with simple linear perturbation theory once one
specifies the baryon-to-photon ratio and matter-radiation ratio, both of
which can be measured from the details of the CMB acoustic peaks
\citep{BenTurWhi97,Hu97ss,HuDod02,WhiCoh02}.
With this scale in hand, one can measure the angular diameter distance and
Hubble parameter as functions of redshift using large galaxy redshift surveys.
This standard ruler offers a robust route to the study of dark energy.

However, the clustering of galaxies does not exactly recover the linear-theory
clustering pattern.  Non-linear gravitational collapse, galaxy clustering bias, and
redshift distortions all modify galaxy clustering relative to that of the
linear-regime matter correlations.  In particular, simulations have shown that
non-linear structure formation and, to a lesser extent, redshift distortions
erase the higher harmonics of the acoustic oscillations
\citep{Mei99,Seo05,Spr05,Whi05}.
This degrades the measurement of the acoustic scale.  More worrisome
is the possibility that these effects might actually bias the measurement.

In this paper, we present a physical explanation of these non-linear
distortions.  We use a hybrid of analytic and numerical methods to
produce a quantitative model for the degradation.  We also investigate
whether the non-linearities can measurably shift the acoustic scale.
The results offer the opportunity to calibrate the effects of 
non-linearities with simulations far smaller than what would be
required to see the effects directly in the acoustic signature.

We organize the paper as following.  
In \S~\ref{sec:review}, we present a pedagogical review of the acoustic
phenomenon, explaining the effects in both the configuration and Fourier bases.
In \S~\ref{sec:nonlin}, we discuss how non-linearities enter the process and
why the large preferred scale of the acoustic oscillations provides an
important simplification of the dynamics.
In \S~\ref{sec:lagrange}, we present a quantitative framework for understanding
the non-linear effects based on Lagrangian displacements.
We construct Zel'dovich approximation estimates of the displacements
in \S~\ref{sec:analytic} and then measure the required distributions
numerically and compare them to the non-linear two-point functions
in \S~\ref{sec:numerical}.
The model is extended to biased tracers in \S~\ref{sec:biased}.
In \S~\ref{sec:shift}, we study how non-linearities could create
systematic shifts in the acoustic scale.
We conclude in \S~\ref{sec:conclusions}.

Throughout the paper, we will refer to redshift space as the position
of galaxies or matter inferred from redshifts, with no correction for peculiar
velocities.  The true position of the galaxies or matter will be 
called real space.  Sometimes, the term real space is used in cosmology
as the complement to the Fourier basis; we will instead refer to these
as Fourier and configuration space (or basis) to avoid confusion.
We use comoving distance units in all cases.

\section{Acoustic oscillations in configuration and Fourier space}
\label{sec:review}

This section is pedagogical in nature.  The physics of the acoustic
oscillations have been well studied in Fourier space,
but we have found that the story in configuration space is rarely
discussed even though it exhibits some of the most important features for
our purposes.  Moreover, this pedagogy can warm up the reader to the
configuration-space view of the non-linearities that we pursue in the rest
of the paper.

The epoch of recombination marks the time when the Universe has cooled enough
that protons and electrons combine to form hydrogen \citep{Pee68,ZKS69,SSS99}.
Prior to this time, the mean free path of the photons against scattering off
of the free electrons is much less than the Hubble distance.
This means that gravitational forces attempting to compress the plasma must
also increase the photon density.  This produces an increase in temperature
and hence in radiation pressure.
Any perturbation in the baryon-photon plasma thus behaves as an acoustic wave.

\subsection{Fourier space}

Before looking at this picture in configuration space, let us remind
ourselves of the physics in Fourier space.  Consider a (standing) plane
wave perturbation of comoving wavenumber $k$ in Newtonian gauge
\citep[e.g.][]{KodSas}.
At early times, the density is high and the scattering is rapid compared with
the travel time across a wavelength.  We may therefore expand the momentum
conservation (Euler) equation in powers of the Compton mean free path over the
wavelength $k/\dot{\tau}$, where $\dot\tau$ is the differential
Compton optical depth and overdots denote derivatives with respect to 
conformal time $\eta = \int dt/a$.
To lowest order and setting $c=1$, we obtain the
{\it tight coupling\/} approximation for the evolution of a 
single Fourier mode of the baryon density perturbation
[\citet{PeeYu70,DZS78} or for a more recent derivation in the relevant gauge
and notation \citet{HuWhi96}]
\begin{equation}
{d \over d\eta}\left[(1+R)\dot\delta_b\right] + {k^2 \over 3}\delta_b
  = -k^2(1+R)\Psi - {d \over d\eta}\left[3(1+R)\dot\Phi\right].
\end{equation}
Here 
$R\equiv 3\rho_b/4\rho_\gamma$ is the baryon-to-photon momentum density ratio
and the gravitational sources are $\Phi$, the perturbation to the spatial
curvature, and $\Psi$, the Newtonian potential.  
At late times, the gravitational perturbations are
dominated by the CDM and baryons (if $\Omega_b/\Omega_m$ is large), while
at early times they are dominated by the relativistic species.
We recognize this equation as that of a driven oscillator with natural
frequency $c_s k$, where the speed of sound $c_s$ is $c/\sqrt{3(1+R)}$.
During the tight coupling phase, the amplitude of the baryon perturbations
cannot grow and instead undergoes harmonic motion with an amplitude that decays
as $(1+R)^{-1/4}$ and a velocity that decays as $(1+R)^{-3/4}$
\citep{HuSug95,HuWhi96}.
For currently accepted values of $\Omega_b$ \citep{Spe06,Ste06}, 
$R<0.3$ at $z>1000$ so the decay of the period is small.


If we define the optical depth 
$\tau_b(\eta)\equiv\int_\eta^{\rm today}\ d\eta\;\dot\tau/(1+R)$, 
we find that the baryons decouple from the photons when $\tau_b\sim1$,
a time we shall refer to as ``decoupling''.  The oscillations in the
baryons are frozen in at this epoch, which occurs after recombination
for currently accepted values of $\Omega_b$.  Expanding to higher order
in $k/\dot{\tau}$, one finds the oscillations are exponentially damped
to due photon-baryon diffusion \citep{Sil68} with a characteristic scale
which is approximately the geometric mean of the horizon and the mean
free path at decoupling (of order $10$~Mpc for a concordance cosmology).
Including the finite duration of recombination slightly alters the damping
from this exponential form \citep{HuWhi97}, but this detail will not be
of relevance to us here.  Note that since decoupling occurs slightly
after recombination, both the sound horizon and the damping scale are
slightly larger than the corresponding scales in the CMB.

Once the photons release their hold on the baryons, the latter can be treated
as pressureless allowing the perturbations to grow by gravitational instability
(e.g.~proportionally to the scale factor in a matter-dominated Universe).
The density and
velocity perturbations from the tight-coupling era, including
the CDM perturbations, must be matched onto the
growing and decaying modes for the pressureless components.
The final spectrum is the component which projects onto the
growing mode.
As is well known \citep{Sun70,PreVis80} --- see \citet{Pad93}, \S8.2 for
a textbook treatment --- at high $k$ the growing mode is sourced primarily
by the velocities, while at low $k$ it is a mixture of density and velocity
terms.  For this reason, at high $k$ the
oscillations in the mass power spectrum are out of phase with the peaks in
the CMB anisotropy spectrum, which arise predominantly from photon densities.
In this limit, extrema occur at $ks_*=(2j+1)\pi/2$
where $j=0,1,2,\ldots$ and
$s_*$ is the sound horizon at decoupling: $s_*=\int c_s\, d\eta$.  In detail,
this differs from the sound horizon at recombination (which controls the
position of the CMB peaks) but the two horizons are comparable.
The amplitude of the oscillations depends on both the driving force ($\Phi$
and $\Psi$) and on $R$.
Since the potentials decay in the radiation-dominated epoch, larger
oscillations come from higher $\Omega_b h^2$ and lower $\Omega_m h^2$
\citep{Eis98,Mei99}.

\subsection{Configuration space}

\begin{figure}[p]
\centerline{\epsfxsize=2.3in\epsffile{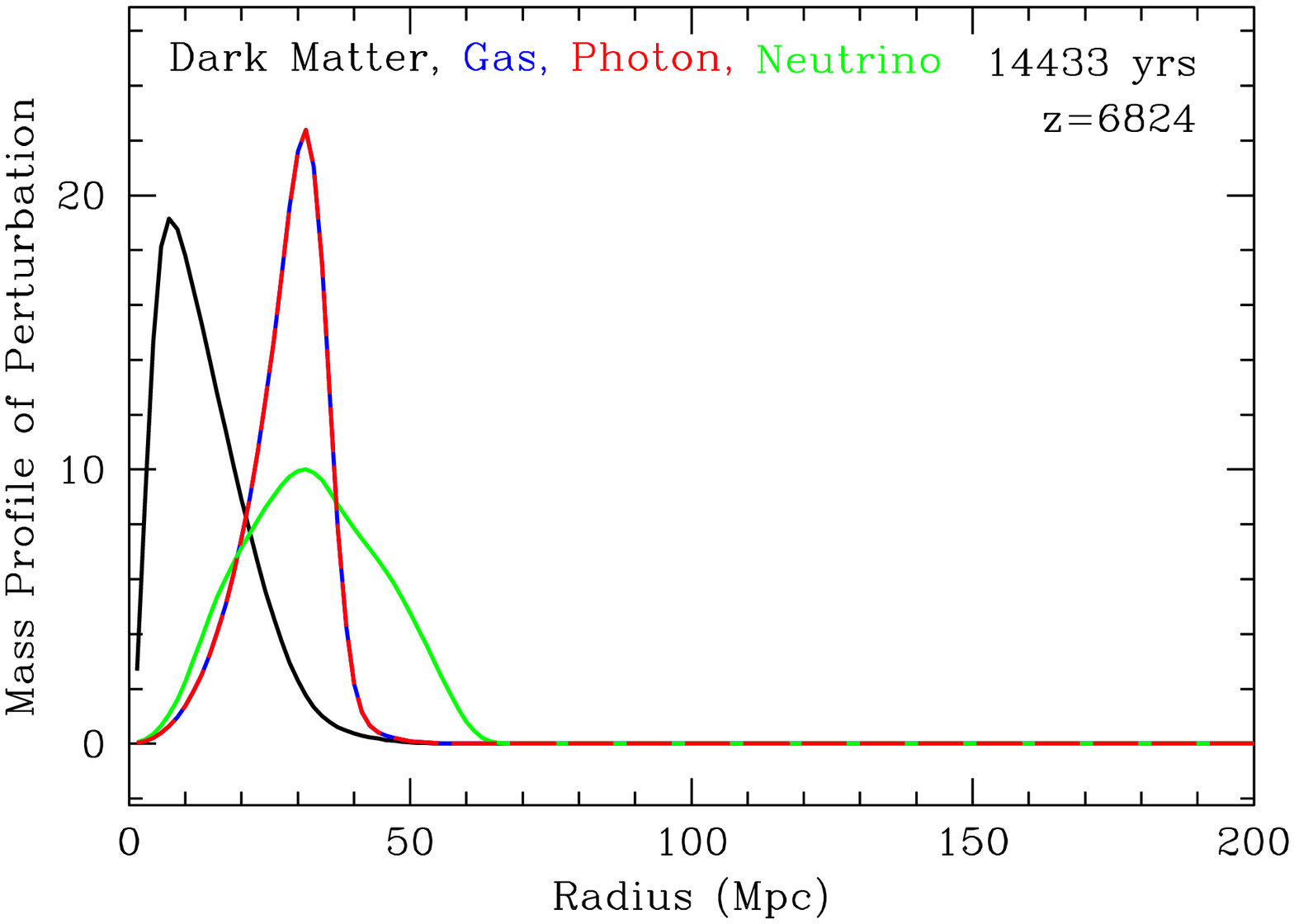}
    \quad   \epsfxsize=2.3in\epsffile{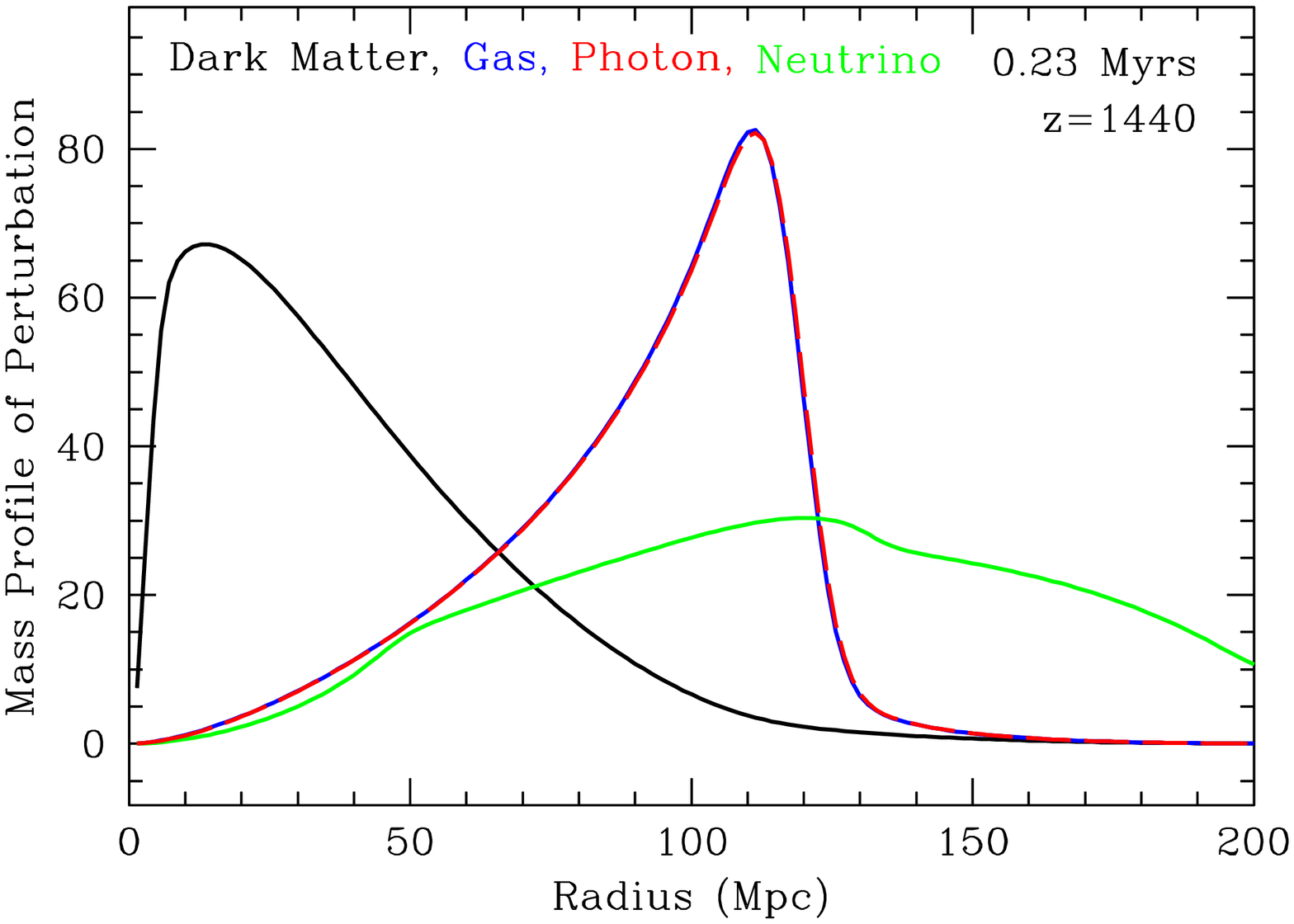}}
\centerline{\epsfxsize=2.3in\epsffile{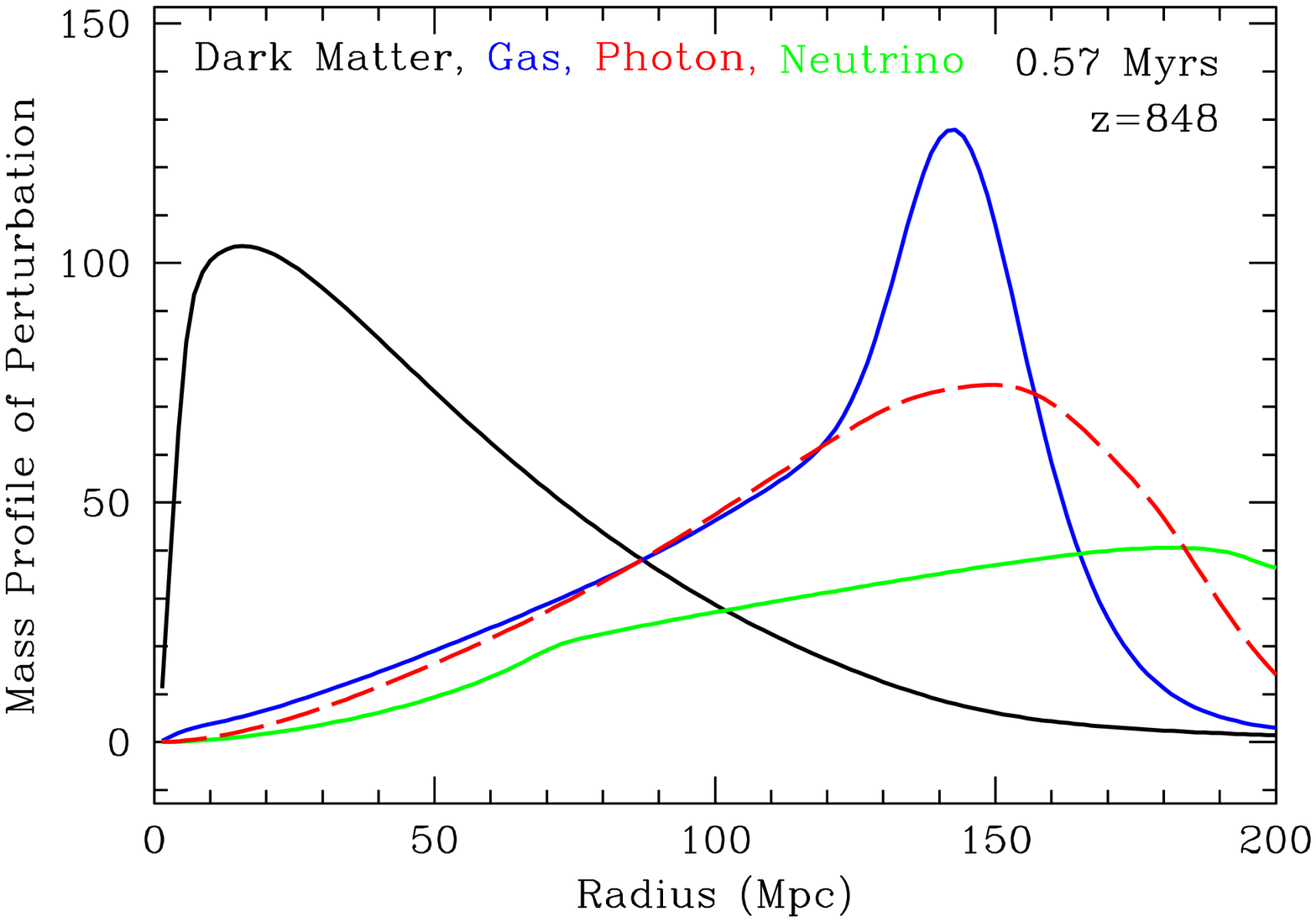}
    \quad   \epsfxsize=2.3in\epsffile{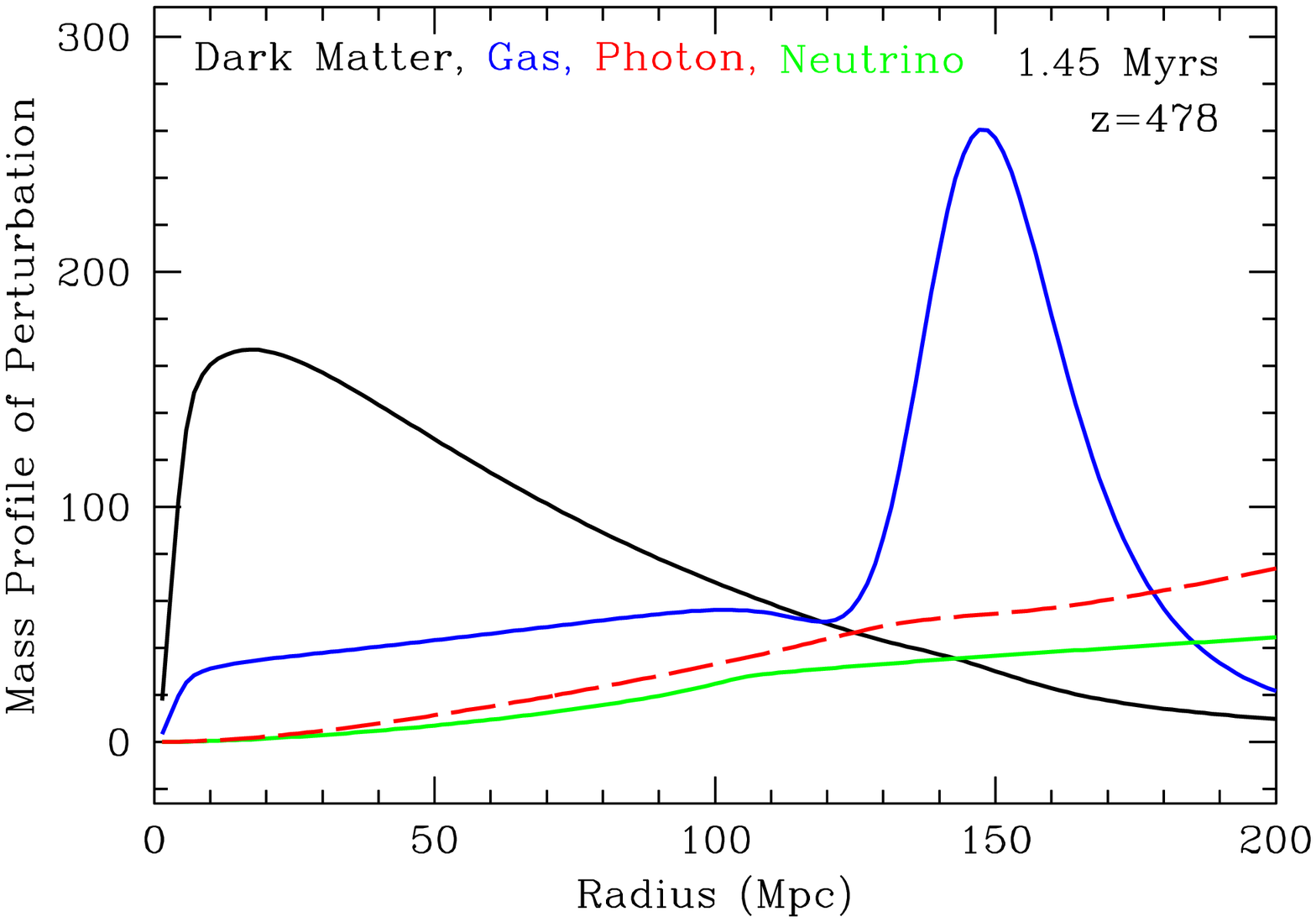}}
\centerline{\epsfxsize=2.3in\epsffile{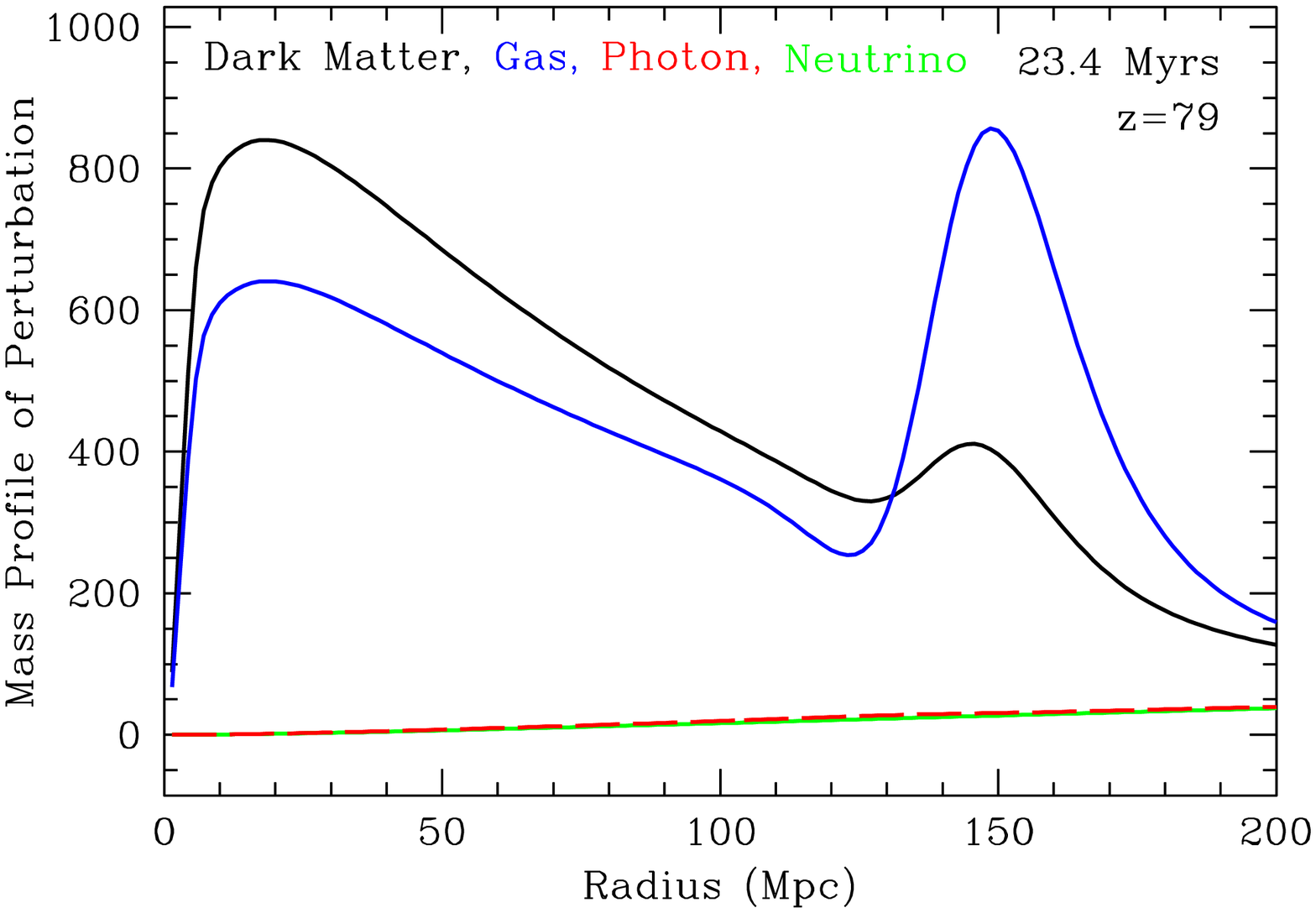}
    \quad   \epsfxsize=2.3in\epsffile{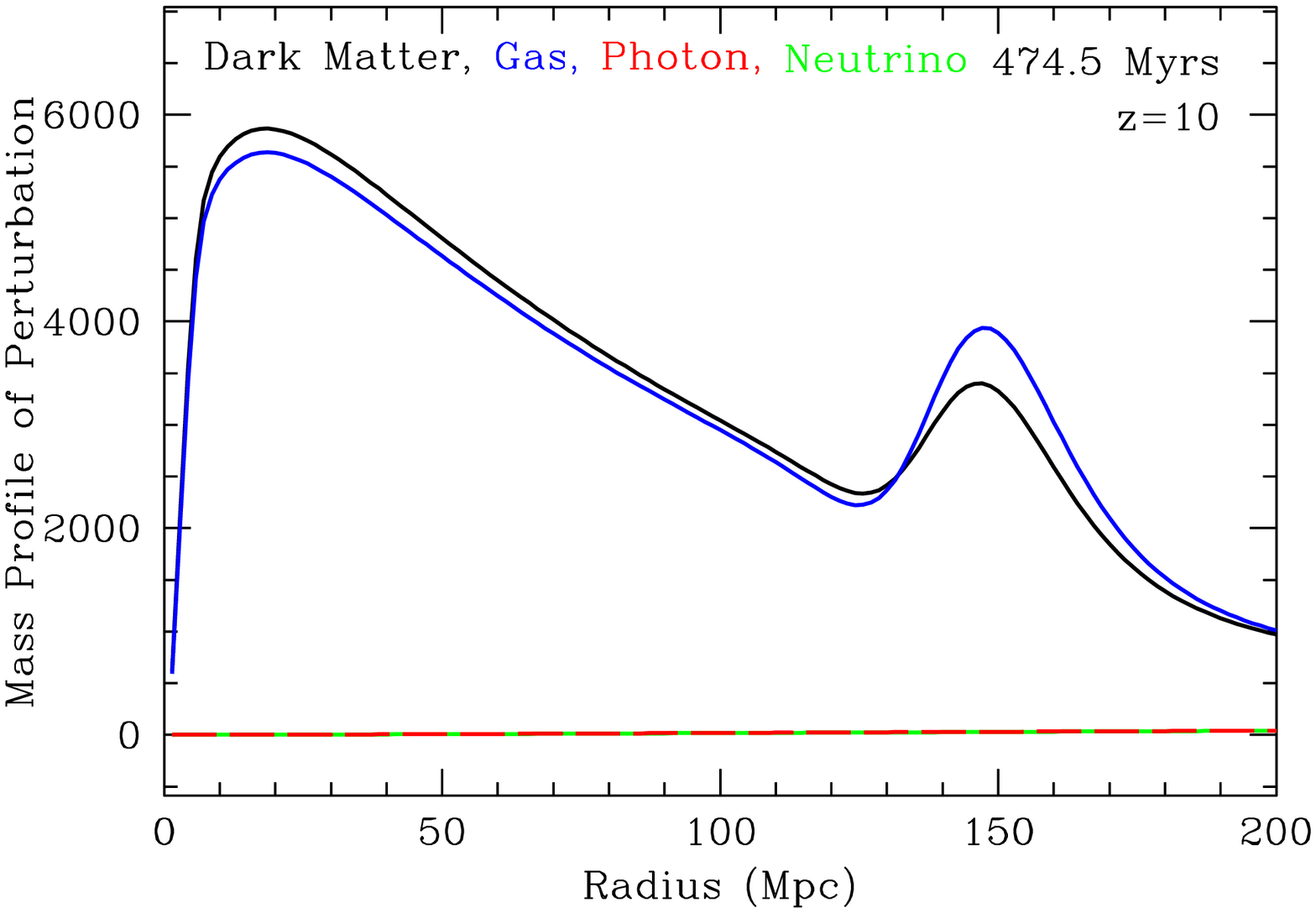}}
\caption{\label{fig:Tevol}
Snapshots of evolution of the radial mass profile versus comoving
radius of an initially point-like overdensity located at the 
origin.  All perturbations are fractional for that species;
moreover, the relativistic species have had their energy density
perturbation divided by 4/3 to put them on the same scale.
The black, blue, red, and green lines are the dark matter,
baryons, photons, and neutrinos, respectively.  The redshift
and time after the Big Bang are given in each panel.
The units of the mass profile are arbitrary but are correctly
scaled between the panels for the synchronous gauge.
a) Near the initial time, the photons and baryons travel outwards
as a pulse.
b) Approaching recombination, one can see the wake in the cold
dark matter raised by the outward going pulse of baryons and 
relativistic species.
c) At recombination, the photons leak away from the baryonic perturbation.
d) With recombination complete, we are left with a CDM perturbation
towards the center and a baryonic perturbation in a shell.
e) Gravitational instability now takes over, and new baryons and dark matter
are attracted to the overdensities.  
f) At late times, the baryonic fraction of the perturbation is near the
cosmic value, because all of the new material was at the cosmic mean.
These figures were made by suitable transforms of the transfer functions
created by CMBfast \protect\citep{Sel96z,Zal00}.
}
\end{figure}

With this in mind, it is instructive to switch to configuration
space and consider what happens to a point-like initial 
overdensity\footnote{The development of the transfer function in 
configuration space was described in detail by 
\citet{Bas01,Bas02}, although that work was based on a 
planar initial overdensity rather than a point-like one.}.
In an adiabatic model, the overdensity is present in all species
\citep{PeeYu70}.
In particular, because the region is overdense in photons, it is overpressured
relative to its surroundings.
This overpressure must equilibrate by driving a spherical sound wave out into
the baryon-photon plasma.  The cold dark matter perturbation is left at small
radius.  The sound wave travels out at the speed of sound.
At the time of decoupling, the wave stalls as the pressure-supplying photons 
escape and the sound speed plummets.  One ends up with a CDM overdensity
at the center and a baryon overdensity in a spherical shell 150 comoving Mpc
in radius for the concordance cosmology.  At $z\ll 10^3$, both of these
overdensities attract gas and CDM to them, seeding the usual gravitational
instability \citep{PeeYu70}. 
The fraction of gas to CDM approaches the cosmic average as the 
perturbation grows by a factor approaching $10^3$, so that one ends up
with an overdensity of all matter at the center and a spherical echo
at 150 Mpc radius.  Galaxies are more likely to form in these overdensities.
The radius of the sphere marks a preferred separation of galaxies,
which we quantify as a peak in the correlation function on this scale.
This evolution is shown graphically in Figure \ref{fig:Tevol}.

The Universe is of course a superposition of these point-like perturbations,
but as the perturbation theory is exquisitely linear at high redshift, we can 
simply add the solutions.  
The width of the acoustic peak is set by three factors:
Silk damping due to photons leaking out of the sound wave \citep{Sil68},
adiabatic broadening of the wave as the sound speed changes due to the
increasing inertia of the baryons relative to the photons, and the
correlations of the initial perturbations.
In practice, the acoustic peak is about 30 Mpc full width at half maximum
for the concordance cosmology.

There are some other interesting aspects of the physics of this
epoch that are worth mentioning in the configuration-space picture.
First is that the outgoing wave does not actually stop at $z\sim 10^3$
but instead slows around $z\sim 500$.  This is partially due to the 
fact that decoupling is not coincident with recombination but is also
because the coupling to the growing mode is actually dominated by the
velocity field, rather than the density field, at $z\sim 10^3$
\citep{Sun70,PreVis80}.
In other words, the compressing velocity field in front of the wave
actually keys the instability at later time.  

Two other aspects of Figure \ref{fig:Tevol} that may be surprising at
first glance are that the outgoing pulse of neutrino overdensity doesn't
actually remain as a  delta function, as one might expect for a population
traveling radially outward at the speed of light, and that the CDM
perturbation doesn't remain at the origin, as one would expect for a cold
species.  Both of these effects are due to a hidden assumption in the initial 
conditions: although the density field is homogeneous everywhere but the origin,
the velocity field cannot be for a growing mode.
To keep the bulk of the universe homogeneous while growing a perturbation
at the origin matter must be accelerating towards the center; this
acceleration is supplied by the gravitational force from the central
overdensity.
However, in the radiation-dominated epoch the outward going pulse of neutrinos
and photons is carrying away most of the energy density  of the central
perturbation.
This outward going pulse decreases the acceleration, causing the inward flow
of the homogeneous bulk to deviate from the divergenceless flow and generating
the behavior of the CDM and neutrinos mentioned above.
Essentially the outgoing shells of neutrinos and photons raise a wake in the
homogeneous distribution of CDM away from the origin of the perturbation.

The smoothing of the CDM overdensity from a delta function at the
origin is the famous small-scale damping of the CDM power spectrum
in the radiation-dominated epoch
\citep{Lif46,PeeYu70,Sun70,GroPee75,DZS78,WilSil81,Pee81,Blu84,BonEfs84}.
The overdensity raised decreases as a function of radius because the
radiation is decreasing in energy density relative to the inertia of
the CDM; in the matter-dominated regime, the outward-going radiation
has no further effect.  A Universe with more radiation causes a larger
effect that extends to larger radii; this corresponds to the shift in
the CDM power spectrum with the matter-to-radiation ratio.

Returning to the major conceptual point, that of the shell of overdensity
left at the sound horizon, we see immediately that the sound horizon provides
a standard ruler.  The radius of the shell depends simply on the sound speed
and the amount of propagation time \citep{PeeYu70,Sun70,DZS78}.
The sound speed is set by the balance of radiation pressure and inertia
from the baryons; this is controlled simply by the baryon-to-photon 
ratio, which is $\Omega_bh^2$.  The propagation time depends on the
expansion rate in the matter-dominated and radiation-dominated 
regimes; this depends on the redshift of matter-radiation equality,
which depends only on $\Omega_m h^2$ for the standard assumption of
the radiation density (i.e., the standard cosmic neutrino and
photon backgrounds and nothing else).

\subsection{Connecting Fourier \& Configuration space}

Next, it is useful to connect this picture to the Fourier space view
of the evolution of plane waves.
Here we consider not standing waves, where the accumulated phase change
until decoupling determines the positions of the peaks in the power
spectrum, but traveling waves.
We imagine each overdense crest of the initial perturbation launches a planar
sound wave that travels out for 150 Mpc \citep{Bas01,Bas02}.
If this places the baryon perturbation on top of another crest in the initial
wave (now marked by the CDM), then one will get constructive interference in
the density field and a peak in the power spectrum.
If the sound wave lands in the trough of the initial wave, then one gets
a destructive interference.  Hence, one gets a harmonic pattern of
oscillations, with a damping set by the width of the post-recombination
baryon and CDM perturbations from a thin initial plane
(i.e.~the 1-d Green's function).

In our opinion, for understanding the late-time matter spectrum, in particular
the standard ruler aspect, the traveling wave view has considerable pedagogical
advantages.
The fact that the baryon signature in the correlation function has a single
peak is a provocative indication that the physics is easily viewed as the
causal propagation of a signal.
Mathematically, the connection is that the correlation function 
and power spectrum are Fourier transform pairs.  The transform
of a single peak is a harmonic sequence, and the profile of the peak
gives the damping envelope of the oscillations.
We shall draw upon this picture as we consider non-linear evolution in the
following sections.

\section{The non-linear degradation of the acoustic signature}
\label{sec:nonlin}

At low redshift, non-linear structure formation, redshift distortions, and
galaxy clustering bias all act to degrade the acoustic signature in the
galaxy power spectrum \citep{Mei99,Seo05,Spr05,Whi05}.
These effects reduce the amount of distance-scale information that can be
extracted from a survey of a given volume \citep{Seo03,Bla06}.
In principle, they might also shift the scale, resulting in biased
information from uncorrected estimators.
One should of course distinguish such shifts from a simple increase in noise;
a noisy statistical estimator is not the same as a biased one.

How do the breakdowns of the ideal linear case enter?  We begin
in Fourier space.  Here the acoustic signature is a damping harmonic
series of peaks and troughs.  As non-linearities progress, mode-mode
coupling begins to alter the power spectrum.  Power increases sharply
at large wavenumber, but the mode-mode coupling also damps out 
the initial power spectrum on intermediate scales, decreasing the 
acoustic oscillations \citep{Gor86,Jai94,Mei99,Jeo06}.
Both of these effects reduce the contrast of the oscillations, making
them harder to measure.

At $z<1$, N-body simulations and galaxy catalogs derived therefrom
show an excess of power at the location of the higher acoustic 
harmonics ($k\approx0.2\ihmpc$).  Clearly the non-linear power
spectrum deviates from the linear one at wavenumbers containing
acoustic information.  Does this mean that one cannot recover
``linear'' information from these scales?  As we'll see next, the
answer is not this pessimistic.

In configuration space, the acoustic signature is a single peak in the
correlation function at $150\mpc$ separation.  The width of the peak is
related to the decreasing envelope of the oscillations in the power spectrum.
The action of the non-linear effects is easy to see: velocity flows
and non-linear collapse move matter in the Universe around by of order
$10\mpc$ relative to its initial comoving position.  This tends to
move pairs out of the $150\mpc$ peak, broadening it.  This broadening
corresponds to the decay of the high harmonics in $P(k)$.  One also
sees that because redshift distortions further increase the distance
between a galaxy's measured position and its initial position,
the acoustic oscillations are further reduced in redshift space.

The key insight from the configuration-space picture is that the processes
that are erasing the acoustic signature are {\it not} at the fundamental
scale of $150\mpc$ but are at the cluster-formation scale, a factor
of 10 smaller.  This is very different from the apparent behavior
in Fourier space that suggests the non-linearities are encroaching on the 
acoustic scale.  The solution to the paradox is to note that the 
acoustic information in Fourier space is actually carried by the 
comparison of adjacent peaks and troughs.
The beat frequency between these nearby wavenumbers is much smaller,
about $\Delta k=0.03\ihmpc$, and is well below the non-linear wavenumber.
In other language, the power that is being built by mode-mode
coupling must be smooth on the scale of $\Delta k$ because the 
non-linearities are developing on much smaller scales\footnote{The
power that is being removed by mode-mode coupling can have structure
on the beat frequency scale because a simple damping does bring 
out one power of the initial power spectrum \protect\citep[e.g.,][]{Cro05}.}.
The existence of this small beat frequency changes the behavior of
non-linearities relative to our usual intuition, which are based on
broad-band effects.

It is also interesting to note that the power that is being introduced
by halo formation on small scales is particularly smooth at the 
wavenumbers of interest \citep{Sch06}.  In the correlation function,
this non-linearity adds very little to $\xi$ at the acoustic scale; it
is purely an excess correlation on halo scales.  This reinforces
the idea that such power can be removed by marginalizing over
smooth broadband spectra
without biasing the recovery of the
acoustic scale.  The important effect of the small-scale non-linearity
is the increase in noise.  This is clear in the power spectrum, where
the noise even in the Gaussian limit is increased by the extra near-white
noise at small wavenumbers.  It is less obvious in the correlation
function, as $\xi(r)$ itself has not changed on the acoustic scale.
However, the noise will be present in the covariance matrix of the 
correlation function, as the near-white noise acts as extra shot noise
that increases the variance in the correlation function.

As for the acoustic scale itself, in configuration space, it is easy to see
that it will not be shifted much at all.  The scale of $150\mpc$
is far larger than any known non-linear effect in cosmology.
To shift the peak, we need to imagine effects that would systematically
move pairs at $150\mpc$ to smaller or larger separations.
We will return to this point in detail in \S \ref{sec:shift} and argue
that non-linear gravitational physics can only introduce shifts of order
0.5\% even at first order and that linearly biased tracers imply a 
cancellation of the first-order term.

Hence, the picture of the non-linear evolution of the acoustic signature
is easy to see in configuration space: the characteristic separation of $150\mpc$
is not shifted but rather is blurred by the action of cluster formation
and bulk flows that move matter around by $\sim\!10\mpc$.  This smears out
the peak in the correlation function, further damping out the higher
harmonics in the power spectrum.  The acoustic scale becomes harder
to measure in a survey because the wider correlation peak is harder
to centroid or because the higher harmonics are at lower contrast 
in the power spectrum.  The existence of collapsed halos adds extra
near-white noise in the power spectrum and creates increased variance
in measurements of the correlation function.

Again, there is nothing in the configuration-space version of the story
that can't be stated in the Fourier-space version, but one must
remember that the small beat frequency between the peaks and
troughs of the acoustic oscillations means that one's usual intuition
about broadband Fourier effects doesn't fully apply.  In particular,
the fact that the power spectrum does show non-linear alterations
at $k=0.1$--$0.2\ihmpc$ does not mean that the acoustic effects
are hopelessly muddied by non-linearities.

\section{Modeling the degradation as a Lagrangian displacement}
\label{sec:lagrange}

We have argued that the erasure of the acoustic signature can
be understood as being due to the motions of matter and galaxies 
relative to the initial preferred separation.  It is therefore
interesting to think about the displacement of matter or tracers
from their position in the near-homogeneous initial state.  This
is the Lagrangian displacement, well known from the Zel'dovich
approximation \citep{Zel70}.

We write the Lagrangian displacement as a vector $\bfu(\bfr)$,
where $\bfr$ is the comoving coordinate system for the homogeneous
cosmology.  One can then write the correlation function at a 
particular (large) separation as 
\beq\label{eq:conv}
\xi(\bfs) = \int d\bfr_{12}\,d\bfu_{12}
	\,d\delta_1\,d\delta_2\, p(\delta_1,\delta_2)\,
	\delta_1 \delta_2 \,p(\bfu_{12}|\bfr_{12},\delta_1,\delta_2)\,
	\delta^{(3)}(\bfr_{12}+\bfu_{12}-\bfs)
\eeq
where the subscripts 1 and 2 indicate the density values at 
two different points, $\bfr_{12} = \bfr_1-\bfr_2$, and likewise
for $\bfu_{12}$.  The probability distribution $p(\delta_1,\delta_2)$
is the distribution of the initial density field; on large scales, 
there is a small correlation $\xi(\bfr_{12})$ between the densities.
The probability distribution $p(\bfu_{12}|\bfr_{12},\delta_1,\delta_2)$
is the distribution for the Lagrangian displacements.  If one were
considering local clustering bias, then one could replace the product 
$\delta_1 \delta_2$ by a stochastic bias 
$\delta_{g,1} \delta_{g,2} p(\delta_{g,1}|\delta_1) p(\delta_{g,2}|\delta_2)$.

In general, this integral is similar to the general redshift
distortion problem \citep{Ham98,Sco04,Mat04b}
and does not have an easy solution.
In practice, given how large the desired scale is compared to the scale of
cluster formation and likely galaxy physics, we consider a peak-background
split.  For this, we write the density field as a sum of a large-amplitude
small-scale density field $\delta^S$ and a small-amplitude large-scale density
field $\delta^L$.  
These fields are to be thought of as statically independent.
The small-scale fields at the two points will be statistically independent.
We do similarly for the Lagrangian displacement field.
This yields
\begin{eqnarray}
\xi(\bfs) &=& \int d\bfr_{12}\,d\bfu_1^S\,d\bfu_2^S\,d\bfu_{12}^L
	\,d\delta_1^S\,d\delta_2^S 	
	\,d\delta_1^L\,d\delta_2^L\,
	\,p(\delta_1^L,\delta_2^L) \,p(\delta_1^S) p(\delta_2^S) 
	\,(\delta_1^S+\delta_1^L) (\delta_2^S+\delta_2^L) \nonumber\\
&&	p(\bfu_1^S|\delta_1^S) \,p(\bfu_2^S|\delta_2^S)
	\,p(\bfu_{12}^L|\bfr_{12},\delta_1^L,\delta_2^L)
	\,\delta^{(3)}(\bfr_{12}+\bfu_{12}^L+\bfu_1^S-\bfu_2^S-\bfs)
\label{eq:convLS}
\end{eqnarray}
Only the $\delta_1^L\delta_2^L$ cross-term yields a non-zero result.

If we treat the large-scale Lagrangian displacement as a delta function
at $\bfu_{12}^L=0$ for the moment, then this integral reduces simply
to
\beq
\label{eq:xismear}
\xi(\bfs) = \int d\bfr_{12} \xi(\bfr_{12})
	\int d\delta_1^S\,d\bfu_1^S \,p(\delta_1^S) \,p(\bfu_1^S|\delta_1^S)
	\int d\delta_2^S\,d\bfu_2^S \,p(\delta_2^S) \,p(\bfu_2^S|\delta_2^S)
	\,\delta^{(3)}(\bfr_{12}+\bfu_1^S-\bfu_2^S-\bfs).
\eeq
In other words, this is simply the correlation function convolved 
twice with the distribution $p(\bfu^S)$ of small-scale Lagrangian 
displacements.  This function can depend on the galaxy bias model.
In the case of the redshift-space correlation function, one can 
simply treat $p(\bfu^S)$ as the difference between the initial
position (which is the same in real or redshift space) and the final
position in redshift space; the function obviously becomes anisotropic
in this case.  When computing the small-scale 
displacement field $\bfu^S$, one should high-pass filter to remove
bulk flows on scales at or larger than $|\bfs|$.  This correction is important,
because bulk flows are generated on quite large scales in CDM cosmologies.

Of course, the double convolution of the correlation function is a
trivial multiplication in the power spectrum.  Either way, one sees
that the small-scale power has been strongly suppressed.  One should remember
that this procedure only represents the memory of the initial 
correlation pattern; it does not introduce the small-scale non-linear
clustering that builds up the small-scale correlation function or
power spectrum.

We do not have a full solution to the case with an arbitrary large-scale
Lagrangian displacement field.  
However, we find a significant simplification in equation (\ref{eq:conv})
or (\ref{eq:convLS}) if we assume that the distribution of $\bfu_{12}$
or $\bfu_{12}^L$ is independent of $\delta_1$ and $\delta_2$.
This is not true in detail: overdense regions will tend to move
toward one another, underdense regions away. 
The mean $\bfu_{12}$ depends on $\delta_1^L+\delta_2^L$ at linear order,
but as shown in \S~\ref{sec:shift}, the amount of this shift is
much less than the $30\mpc$ width of the acoustic peak.  It is 
also much less than the standard deviation of the $\bfu_{12}$ 
distribution computed in \S~\ref{sec:analytic} or measured in 
\S~\ref{sec:numerical}.
Hence, because the shifts are small and because the density dependence
produces opposing shifts between the overdense and underdense regions, it is a good
approximation to ignore the dependence when computing the broadening
of the acoustic peak.  This would of course not be a good approximation 
for determining whether there is a mean shift in the scale; we
will return to this topic in \S \ref{sec:shift}.


As both the large and small-scale terms have boiled down to the
distribution of displacements unconditioned by the density field,
we suggest that one can handle both of the effects simultaneously
and also do the proper filtering of the large-scale bulk flows by
computing the distribution of $\bfu_{12}$ for pairs of particles
separated by a given $\bfr_{12}$, without regard for the dependence
on $\delta_1$ and $\delta_2$.  Note that this avoids computing
a peak-background split.  The resulting distribution 
$p(\bfu_{12}|\bfr_{12})$ is our model for the 
kernel that the correlation function will be convolved with, 
following equation (\ref{eq:conv}).
We will measure this kernel numerically in \S \ref{sec:numerical} and 
show that it reproduces the non-linear degradation of the 
acoustic signatures in simulated power spectra.

If we consider a local clustering bias \citep{Col93,Sch98,Dek99},
then we can write
\beq
\xi(\bfs) = \int d\bfr_{12}\,d\bfu_{12}
	\,d\delta_1\,d\delta_2\, p(\delta_1,\delta_2)
	\,d\delta_{g,1}\,d\delta_{g,2}\, 
	p(\delta_{g,1}|\delta_1) \,p(\delta_{g,2}|\delta_2)
	\delta_{g,1} \delta_{g,2} \,p(\bfu_{12}|\bfr_{12},\delta_1,\delta_2)
	\,\delta^{(3)}(\bfr_{12}+\bfu_{12}-\bfs)
\eeq
where $\bfu$ refer the displacements of the galaxies.
In the peak-background split, the $\delta_{g,1}$ factor becomes
$\delta_{g,1}^S+b\delta_1^L$, where $b$ is the large-scale bias.
In equation (\ref{eq:xismear}), the terms of the form
\beq
\int d\delta_1^S\,d\bfu_1^S p(\delta_1^S) \,p(\bfu_1^S|\delta_1^S)
\eeq
become
\beq
b \int d\delta_1^S\,d\delta_{g,1}\,d\bfu_1^S 
\,p(\delta_1^S) \,p(\delta_{g,1}^S|\delta_1^S) \,p(\bfu_1^S|\delta_1^S).
\eeq
In a simulation, this reduces to the distribution of $\bfu_1$
for galaxy ``particles''.  
Again, we suggest simply computing $\bfu_1-\bfu_2$
for pairs of galaxies separated by $\bfr_{12}$.  Of course, 
the initial position of a galaxy is ill-defined in a simulation,
but as we are interested in displacements similar in size to the
width of the acoustic peak, we expect that any reasonable measure
of the initial position of the center of mass of the galaxy will agree
to better than 1 Mpc, which is negligible in the scatter of the 
displacements.  It is important to note that the distribution
of the displacements
of the galaxies may be different from that of the mass displacement,
particularly in redshift space.

The picture of a peak and a spherical echo might lead one to hope
that one could pick tracers that would preferentially accentuate
the peak or sphere, therefore boosting the height of the peak in 
the correlation function.  We show explicitly in Appendix \ref{sec:bias}
that this is not possible for local bias, at least within the Gaussian
random field assumption; the large-scale correlation function of
galaxies always tracks that of the matter, as known from theorems
about local bias \citep{Col93,Sch98}.

\section{Analytic Estimates}  \label{sec:analytic}

In the Zel'dovich approximation our interest in the Lagrangian displacements
translates into study of correlations in the initial velocity field at two
points.  The displacement is
\begin{equation}\label{eq:zel}
  \bfu_{12} = \int {d\bfk\over (2\pi)^3} \delta_{\bfk} {\bfk\over ik^2}
  	     \left[ e^{i\bfk\cdot\bfr_1} - e^{i\bfk\cdot\bfr_2} \right].
\end{equation}
This quantity has a mean of zero, but of course the variance is non-zero.
The variance of the displacement along the direction between the
two points is
\beq
\left\langle u_{12,\parallel}^2\right\rangle =
        \int {d\bfk\over (2\pi)^3}
	P(k) \left(\bfk\cdot\bfr_{12}\over k^2 r_{12}\right)^2
	\left|e^{i\bfk\cdot\bfr_{12}}-1\right|^2
    = r_{12}^2 \int {k^2\,dk\over 2\pi^2} P(k) f_\parallel(kr_{12})
\eeq
where
\beq
f_\parallel(x) = {2\over x^2} \left({1\over3}
        - {\sin x\over x} - {2\cos x\over x^2}
	+ {2\sin x\over x^3}\right).
\eeq
Note that $f_{||}\rightarrow 1/5-x^2/84$ as $x\rightarrow0$, so the
result behaves similarly as $k\rightarrow0$ as the integral to 
compute the density fluctuations,
rather than the $k^{-2}$ behavior of the rms bulk flow integrand.
This is because very large wavelength modes move both particles together.
The variance of the displacement in a direction orthogonal to $\bfr_{12}$
is the same formula, but with
\beq
f_\perp(x) = {2\over x^2}
  \left({1\over3} - {\sin x\over x^3} + {\cos x\over x^2}\right),
\eeq
which behaves as $1/15 - x^2/420$ as $x\rightarrow0$.

For a concordance CDM cosmology with $\sigma_8 =  0.85$ (i.e., $z\approx0$), 
we find rms displacements of $9\hmpc$ along the separation direction and
$8\hmpc$ transverse for separations of $100\hmpc$.
These displacements are mostly generated on large scales, with 50\% of the
variance at $k<0.05\ihmpc$ for radial displacements and $k<0.08\ihmpc$ for
transverse.
Rather little of the variance is generated at $k<0.02\ihmpc$; wavelengths
much longer than $100\hmpc$ move both points similarly.  The displacements
scale linearly with $\sigma_8$ (of the matter) and hence are smaller 
at high redshift.

In redshift space, the Lagrangian displacement in the Zel'dovich approximation
is simply increased by a factor of $f=d(\ln D)/d(\ln a)\approx \Omega_m^{0.56}$ 
along the line of sight, because the same velocity that
gives the real-space displacement also alters the position in redshift space 
\citep{Pee80,Kai87}.

\section{Numerical Results for the Matter Distribution}
\label{sec:numerical}

\begin{figure}[tb]
\plottwo{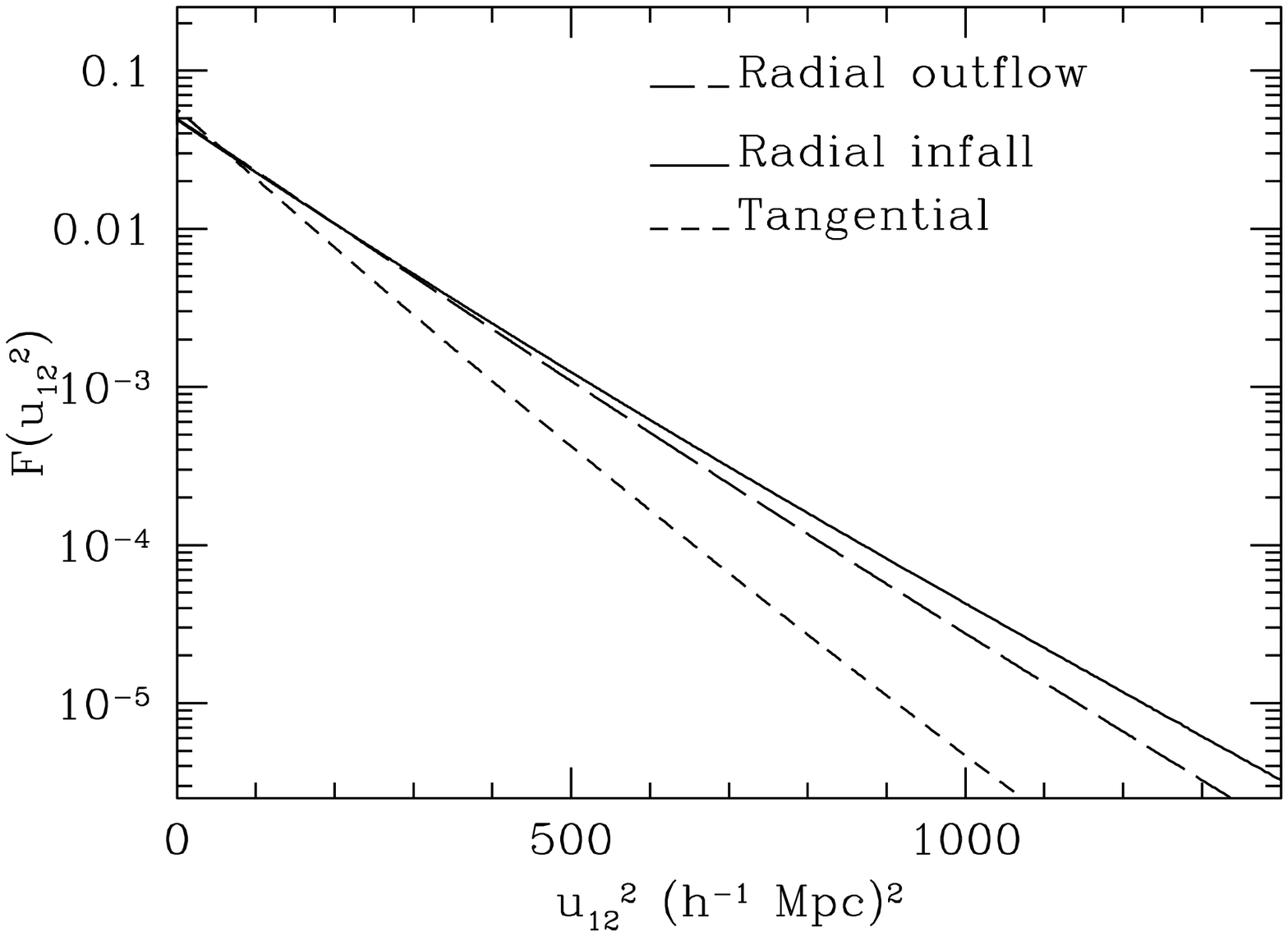}{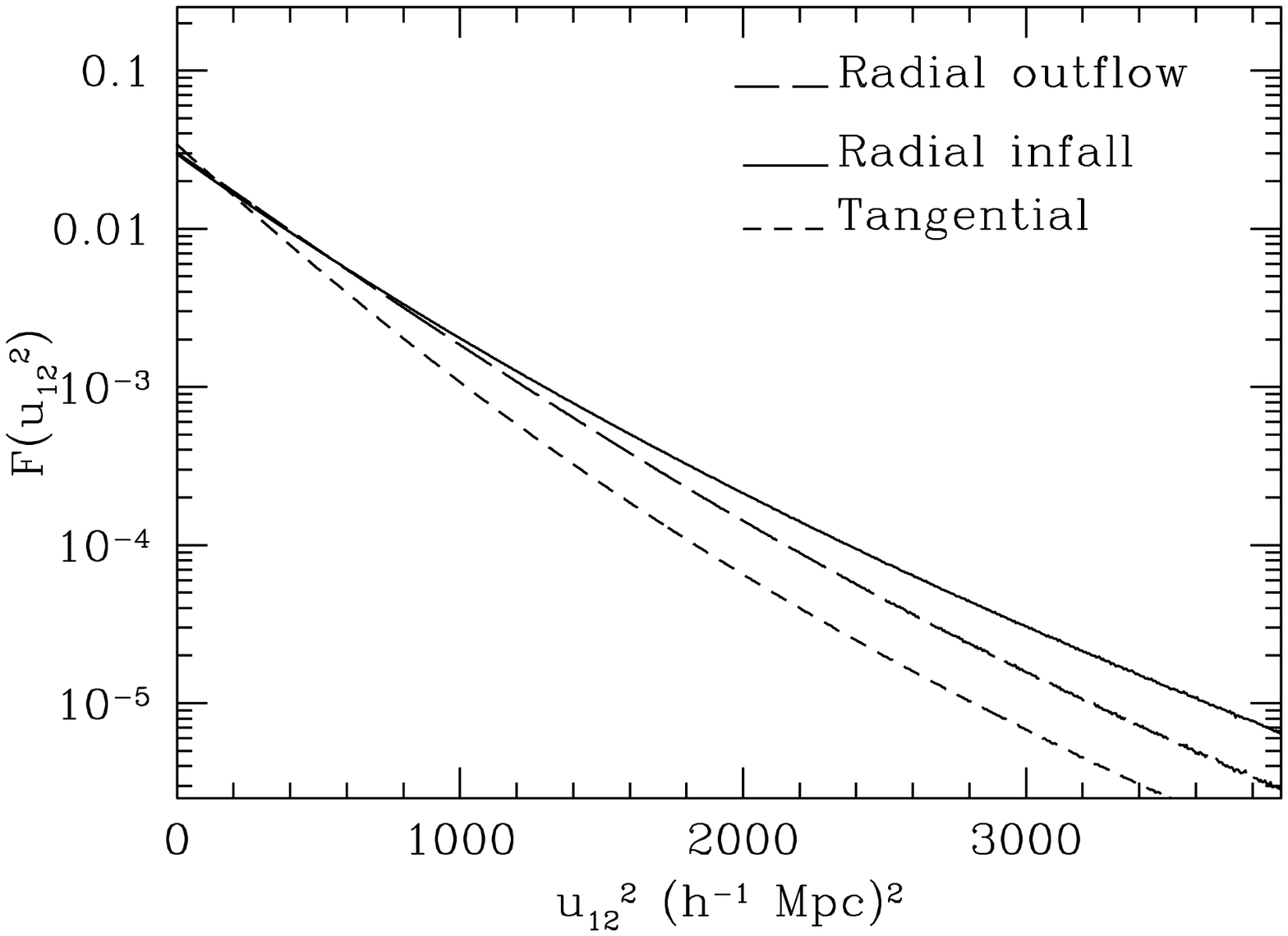}
\caption{\label{fig:distrib}%
The distribution of pairwise Lagrangian displacements for particles
initially separated by $100\hmpc$.  The left panel is in real space;
the right panel is in redshift space.  Both are at redshift $z=0.3$.
The plot is shown as the log of probability versus the square of
the separation so that a Gaussian distribution would be a straight line.
The distribution in the radial direction is slightly skew; we fold 
the distribution at zero and show the infalling and outflowing distribution
as separate lines.
In real space, the distribution is nearly Gaussian; in redshift 
space, it is slightly cuspier.
The displacement in the radial direction has slightly more variance
than that in the tangential direction.  For the redshift-space plot,
the displacement is always in the direction along the line of sight.
The displacements in the direction across the line of sight are of
course identical to those in real space.
}
\end{figure}

\begin{table*}[tb]
\footnotesize
\caption{\label{tab:real}}
\begin{center}
{\sc Lagrangian Displacements Distribution for Matter\\}
\begin{tabular}{rccccccc} 
\doubleline
&    &     & \multicolumn{3}{c}{Radial} & \multicolumn{2}{c}{Transverse} \\
&$z$ & $R$ & $\sigma$ & Skewness & Kurtosis & $\sigma$ & Kurtosis \\
\singleline
Real-space
& 0.3 &  50 & 7.59 & -0.092 & 0.102 &  6.12 & 0.053 \\
& 0.3 &  80 & 8.02 & -0.049 & 0.060 &  6.82 & 0.036 \\
& 0.3 & 100 & 8.15 & -0.034 & 0.049 &  7.10 & 0.029 \\
& 0.3 & 120 & 8.24 & -0.023 & 0.043 &  7.30 & 0.023 \\
& 0.3 & 150 & 8.33 & -0.013 & 0.041 &  7.51 & 0.018 \\ \singleline
& 0.3 & 100 & 8.15 & -0.034 & 0.049 &  7.10 & 0.029 \\
& 0.4 & 100 & 7.76 & -0.032 & 0.047 &  6.77 & 0.026 \\
& 0.5 & 100 & 7.40 & -0.031 & 0.045 &  6.46 & 0.022 \\
& 0.7 & 100 & 6.74 & -0.028 & 0.040 &  5.89 & 0.016 \\
& 1.0 & 100 & 5.90 & -0.024 & 0.035 &  5.17 & 0.005 \\
& 1.5 & 100 & 4.84 & -0.020 & 0.026 &  4.24 & -0.012 \\
& 3.0 & 100 & 3.07 & -0.011 & 0.010 &  2.70 & -0.053 \\  \singleline
%
%
Redshift-space
& 0.3 & 100 & 13.59    & --0.063   & 0.19     & 11.96     & 0.23  \\
& 0.4 & 100 & 13.30    & --0.061   & 0.18     & 11.72     & 0.23  \\
& 0.5 & 100 & 12.98    & --0.059   & 0.18     & 11.45     & 0.22  \\
& 0.7 & 100 & 12.26    & --0.056   & 0.17     & 10.82     & 0.21  \\
& 1.0 & 100 & 11.13    & --0.051   & 0.15     &  9.83     & 0.18  \\
& 1.5 & 100 & 9.41     & --0.044   & 0.11     &  8.30     & 0.14  \\
& 3.0 & 100 & 6.10     & --0.031   & 0.05     &  5.35     & 0.05  \\
\doubleline
\end{tabular}
\end{center}
NOTES.---%
Separations $R$ and rms displacements are given in 
comoving $\hmpc$.  Skewness and kurtosis statistics are the usual
dimensionless normalization.  Negative skewness means that the heavier
tail is inwards. 
The mean displacements are all consistent with round-off error,
less than $10^{-4}R$.
For redshift space, we list the displacements in the line-of-sight
direction when this direction is radial and transverse to the initial
separation vector.  The displacements in the direction perpendicular
to the line-of-sight are the same as in real space.
\end{table*}

We have measured the Lagrangian displacement of pairs of particles from the
set of N-body simulations presented in \citet{Seo05}.  
In brief, we use 30 simulations, each $256^3$ particles in a $512\hmpc$
periodic cube.  The cosmology is $\Omega_m=0.27$, $h=0.72$,
$\Omega_\Lambda =0.73$, $\Omega_b=0.046$, $n=0.99$, and $\sigma_8=0.9$.
We select pairs of particles whose initial separation falls in  a given bin,
e.g., narrowly centered on $100\hmpc$, and compute the difference of their
final separation and initial separation.
We project this vector into components parallel and perpendicular to the
initial separation vector and will refer to these as the radial and transverse
displacements.

In real space, Figure \ref{fig:distrib} shows the distribution of radial and
transverse displacements.  Both distributions have zero mean, as required
by homogeneity, and only mild skewness and kurtosis.  In other words, the
distributions are close to Gaussian, with the radial displacements having
slightly larger variance.  The skewness is such that large inward displacements
are more likely than large outward displacements.
On small scales, many pairs of particles fall into the same halo, so that
$\bfu_{12} = -\bfr_{12}$, and there is necessarily a positive tail that keeps
the mean zero.  In other words, the distribution becomes very skew.
On $100\hmpc$ scales, it appears that treating the distribution as Gaussian
is a good approximation.

The variance of the displacements is hence the relevant statistic to 
quantify the distribution.
We tabulate this in Table \ref{tab:real} at various redshifts and
separations.  The variance is a slow function of
scale: the radial standard deviations at 50, 100, and $150\hmpc$ are
7.59, 8.15, and $8.33\hmpc$, respectively.   We will therefore
focus on $100\hmpc$ separation as a representative value, broadly
applicable around the scales of interest.

\begin{figure}[tb]
\plotone{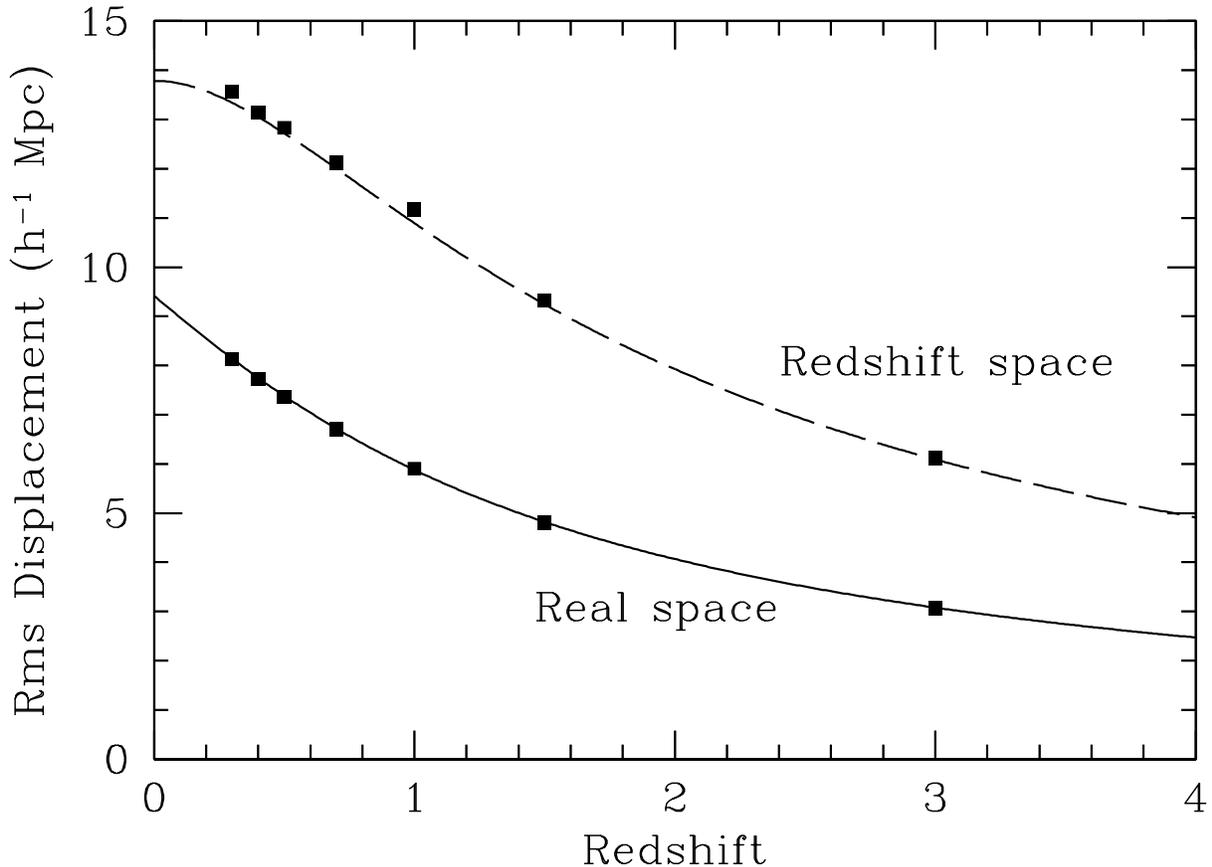}
\caption{\label{fig:disp}%
The rms Lagrangian displacement in the radial direction as a
function of redshift for initial separations of $100\hmpc$.  
The upper points are for redshift space when the initial separation
is along the line of sight.  The lower points are for real space.
The lines are a model in which the real-space result varies as
the growth function $D$, while the redshift-space result varies
as $D(1+f)$, where $f = d\ln D/d\ln a$.  The residuals to the fit
are $\sim\!1\%$, although our simulations
may not be accurate to this level due to their limited mass resolution.
}
\end{figure}

Figure \ref{fig:disp} shows the rms as a function of 
redshift for $100\hmpc$ scale.
The results are well modeled as increasing linearly with the 
growth function; the residuals are below 1\% fractional.  
This is not surprising because so much of
the displacement in the LCDM cosmology is generated by flows
that originate at $k\lesssim0.1\ihmpc$, where the growth is
close to linear theory.  However, it is possible that the
coarseness of simulation resolution may be causing us to 
slightly under-resolve the non-linear behavior at $z=3$, so
the excellence of the growth function scaling might be somewhat
worse in reality.  This does not affect the result for the
acoustic oscillations because the intrinsic width of the 
acoustic peak is broader than the displacement distribution
at that time.

The rms displacements are actually slightly smaller than predicted by
the Zel'dovich approximation.  This is the well-known effect that 
pancake collapse in the Zel'dovich approximation slightly overshoots;
this overshoot prompted development of models such as the adhesion model
\citep{Kof88,Gur89}.  
The slight offset between the radial and tangential rms matches the
prediction of the Zel'dovich approximation.

In redshift space, the displacements along the line of sight are
enhanced.  This is primarily due to the fact that the velocities
and displacements are well correlated on large scales \citep{Kai87}; small-scale
velocity dispersions (e.g., Fingers of God) play a subdominant role.
Again, the displacement distributions are close to Gaussians, with
small skewness.  The kurtosis is larger; the redshift-space distribution 
has larger non-Gaussian tails (Fig.~\ref{fig:disp}).  The means are again zero.

The redshift-space variance is of course a function of the angle
of the displacement vector to the line of sight as well as of
the angle of the displacement to the separation vector.  As 
before, the difference between the radial and transverse 
displacement distributions is small, provided one holds the 
angle of the displacement to the line of sight fixed.
For example, at $z=0.3$ and $100\hmpc$ scales, the radial 
displacement for pairs along the line of sight is $13.59\hmpc$
while the line of sight displacement for pairs transverse to
the line of sight is $11.96\hmpc$.  The displacements in the
direction transverse to the line of sight are of course unchanged
by peculiar velocities.

As it is the radial distribution that most affects the smearing of
the acoustic peak, we will focus on the radial displacement along
and across the line of sight.  The transverse displacement variances
are only slightly smaller, and given their small impact on the
problem anyways, it is a good approximation to simplify by rounding them
up to the radial displacement variances. 
The radial displacement rms is modulated roughly as the square of the
cosine of the angle relative to the line of sight.  
In the direction
along the line of sight, the boost in the standard deviation is very
close to the value $f$ predicted by linear theory.  We show this 
fit in Figure \ref{fig:disp} as well.  The fit appears low by about
1\% fractionally, with about 1\% residual around that correction.

\begin{figure}[p]
\centerline{\epsfxsize=3in\epsffile{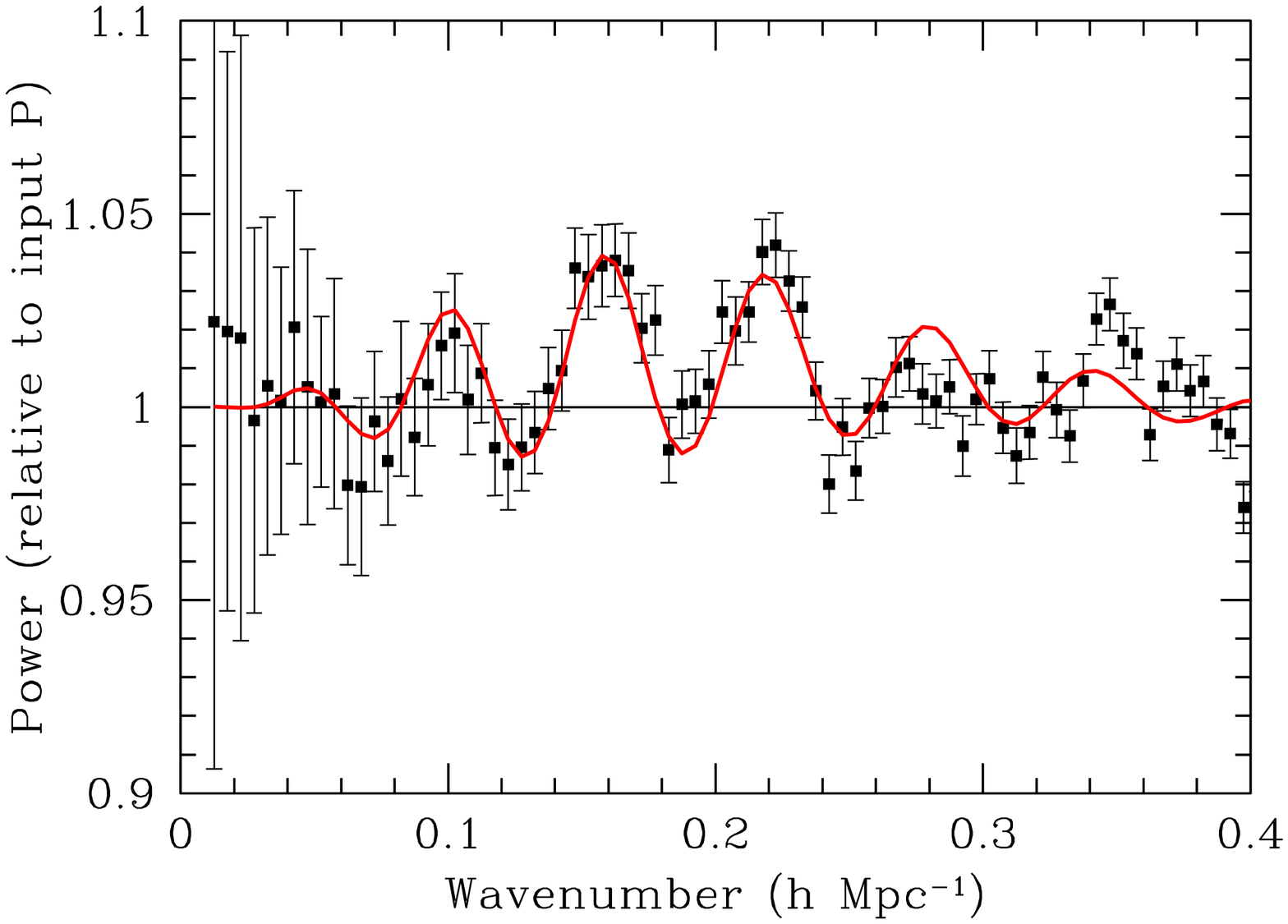}\hspace{\fill}
    \epsfxsize=3in\epsffile{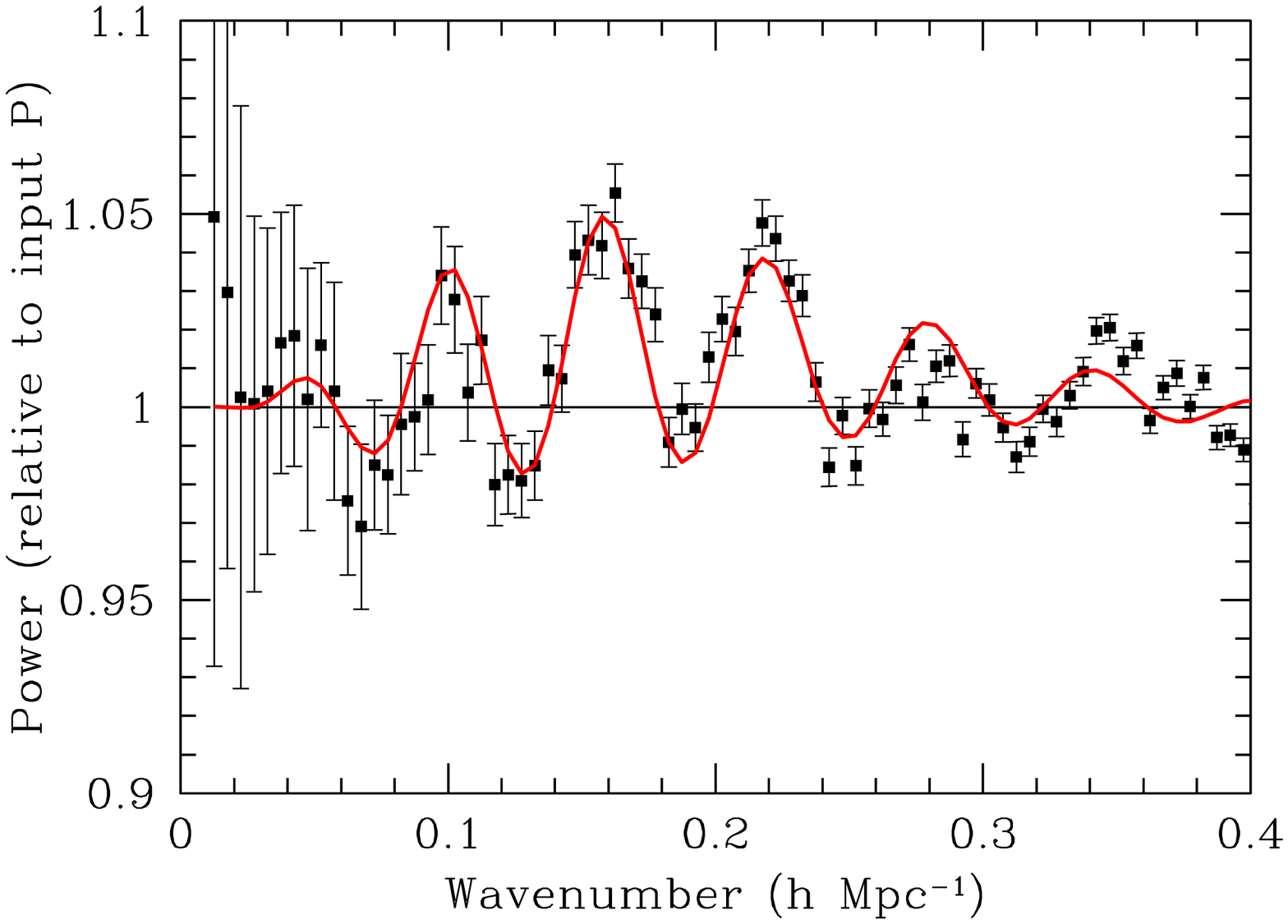}} 
\centerline{\epsfxsize=3in\epsffile{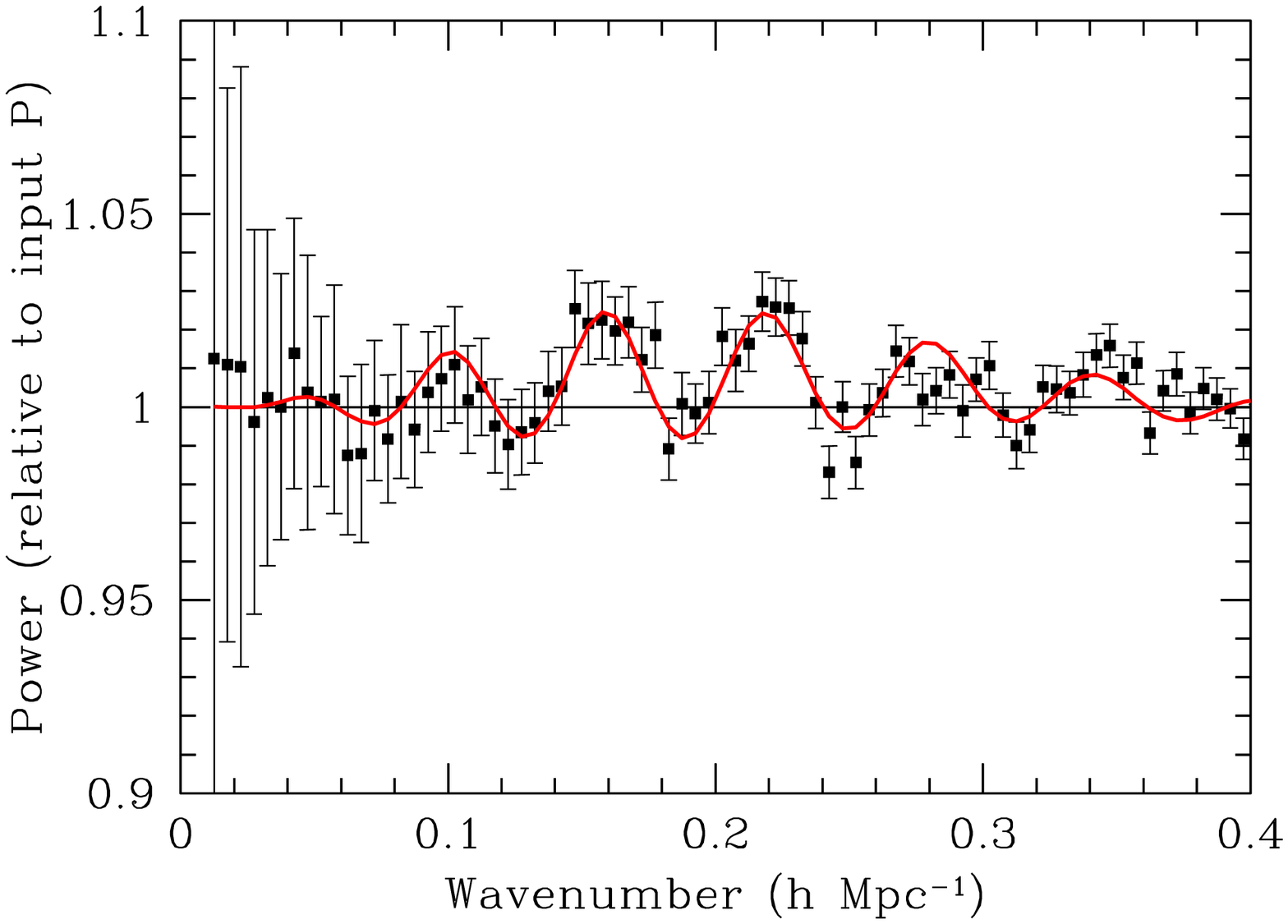}\hspace{\fill}
    \epsfxsize=3in\epsffile{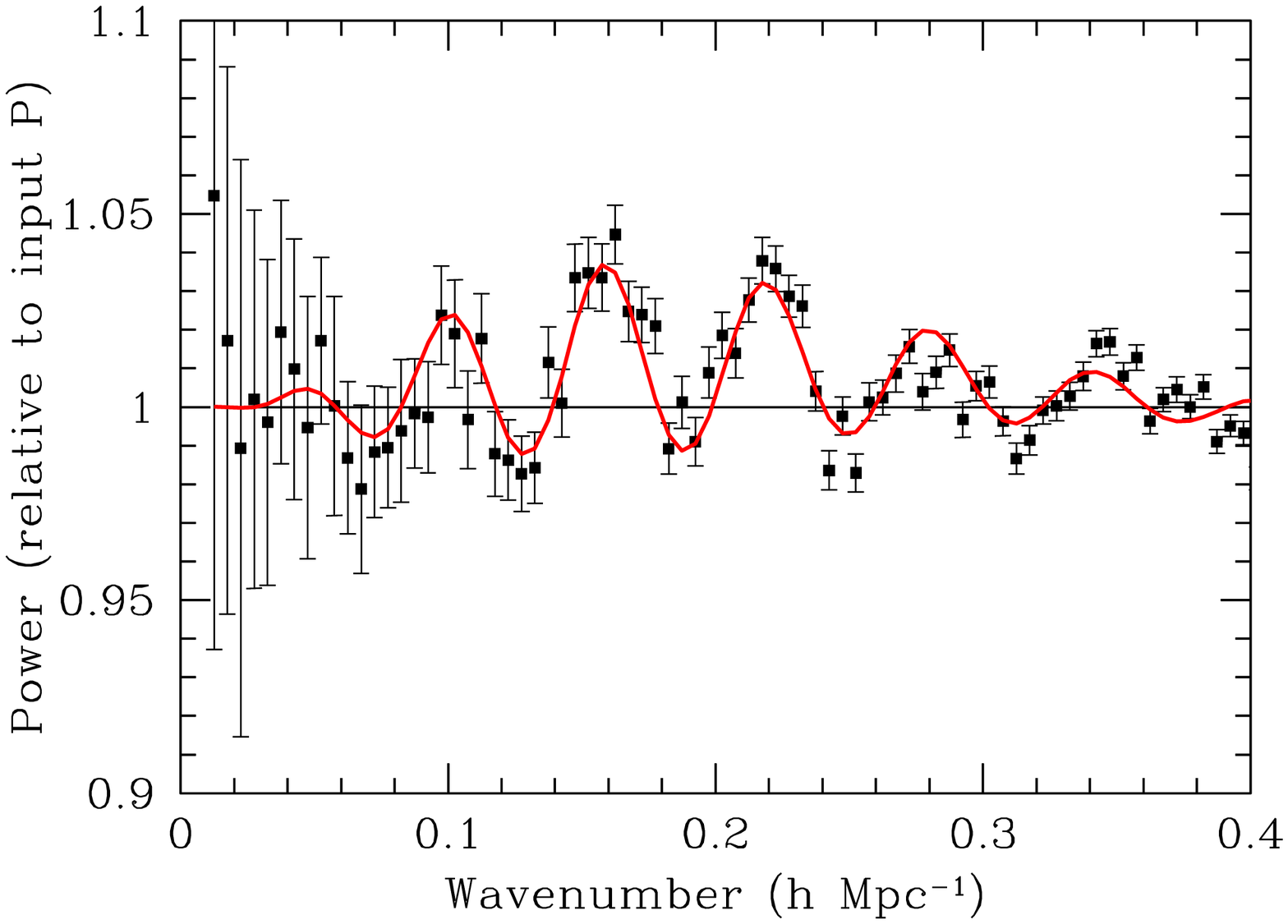}} 
\centerline{\epsfxsize=3in\epsffile{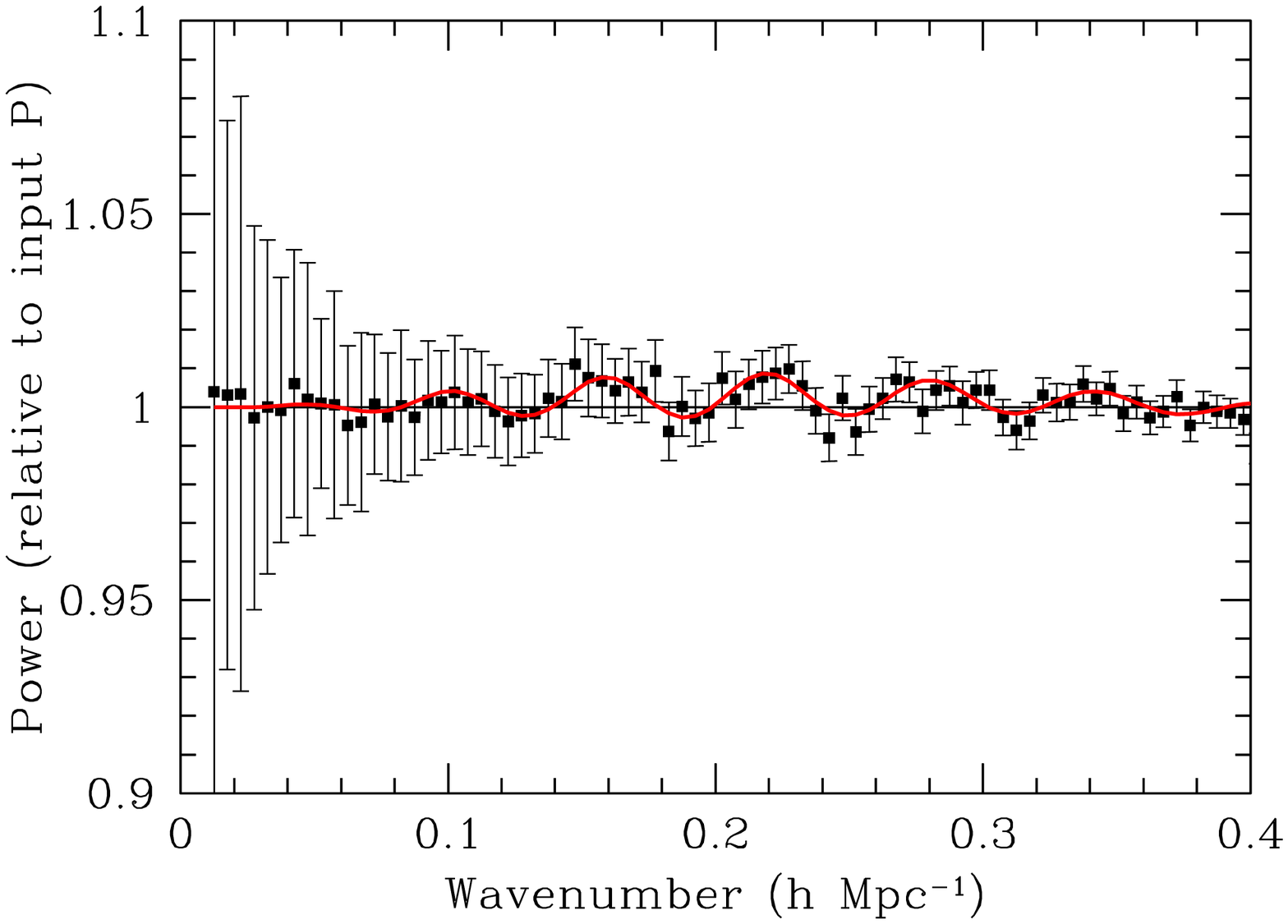}\hspace{\fill}
    \epsfxsize=3in\epsffile{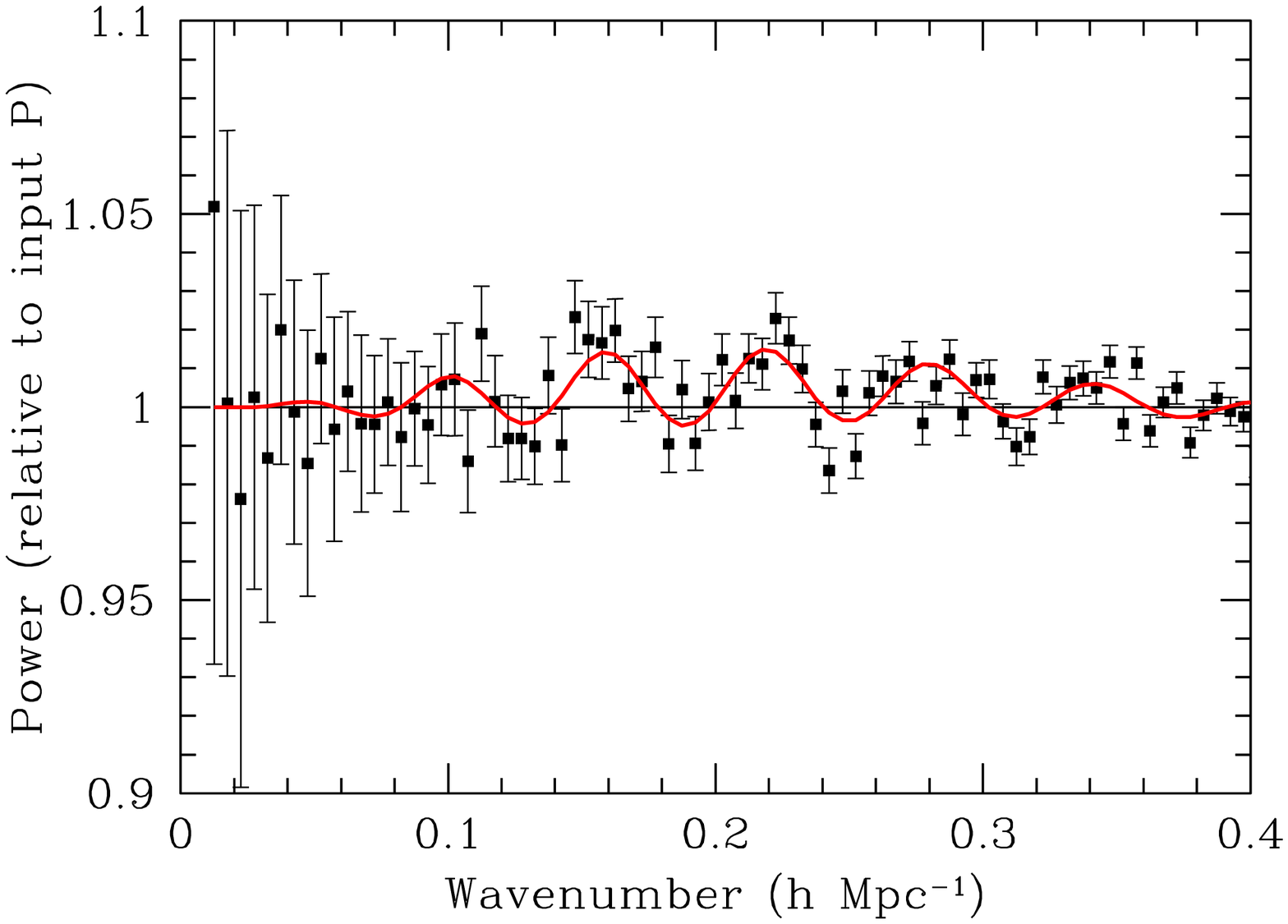}} 
\caption{\label{fig:power}%
The power in the simulations compared to the model.  The left column
shows real space; the right column shows redshift space.
The top row is $z=0.3$, the middle $z=1$, and the bottom row is $z=3$.
In all panels, the simulation power has been fit over the
range $0<k<0.4\ihmpc$ to a model of 
$b^2 P_{\rm smear} + a_0 + a_2 k^2 + a_4 k^4 +a_6 k^6$, where 
$P_{\rm smear}$ is the smeared linear power spectrum $P_{\rm linear}$,
with the small scales restored by the no-wiggle form.
What is plotted is the residual 
$P_{\rm res} = (P_{\rm measured} - a_0-a_2 k^2-a_4 k^4 -a_6 k^6)/b^2$ 
divided by the linear power spectrum.  In detail, we divide by the initial
power spectrum of the simulation, so that the variations in the initial
amplitudes of particular modes cancel out; this doesn't change the sense
of the plot but tightens the residuals slightly.  The smeared model ({\it red}) 
has been similarly divided by the linear power spectrum.  
If the linear spectrum were fully intact, it would be flat in 
this plot.  Erasure of the acoustic peaks produces oscillations
in this plot.  One sees that the model provides an excellent match 
to the observed degradation.
}\end{figure}

In sum, the radial displacements across and along the line of sight 
are well predicted by the following simple model:
\begin{eqnarray}
\sigma_\perp &=& s_0 D \\
\sigma_\parallel &=& s_0 D(1+f)
\end{eqnarray}
where $D$ is the growth function and $f=d(\ln D)/d(\ln a)\approx\Omega^{0.56}$.
The length $s_0$ is $12.4\hmpc$ for $100\hmpc$ separations in our cosmology
if we normalize $D$ such that $D=(1+z)^{-1}$ at high $z$ ($D=0.758$ at $z=0$).
The length will depend slightly on cosmology, at least scaling linearly with
the clustering amplitude (our simulations have a linear $\sigma_8=0.9$ at
$z=0$, but the relevant normalization scale is much above $8\hmpc$).  
$s_0$ is also a slowly increasing function of scale.

If one approximates the transverse displacements as the same,
then the distribution of the displacement vector is very nearly
an elliptical Gaussian, independent of the angle of the separation
vector to the line of sight.  To the extent that one treats $s_0$ as
constant, the effects on the correlation function become a 
simple convolution, meaning that the modification of the power
spectrum is simply to multiply by a Gaussian.  
To be clear, to the extent that the displacement distributions are
Gaussians of rms $\sigma_\parallel$ and $\sigma_\perp$, along and across
the line of sight, we have 
\beq
P(\bfk) = P_{\rm linear}(\bfk)
  \exp\left(-{k_\parallel^2\over 2\sigma_\parallel^2} 
            -{k_\perp^2    \over 2\sigma_\perp^2}\right)
\eeq
This is the model for the portion of the linear power spectrum,
with acoustic peak, that survives.  
Of course, this is a poor model on small scales, as one has filtered out
all of the small-scale clustering.  One can restore the small-scale
linear spectrum, without wiggles, by adding the no-wiggle approximation from
\citet{Eis98} multiplied by unity minus the smearing function.

\begin{figure}[tb]
\centerline{\epsfxsize=3in\epsffile{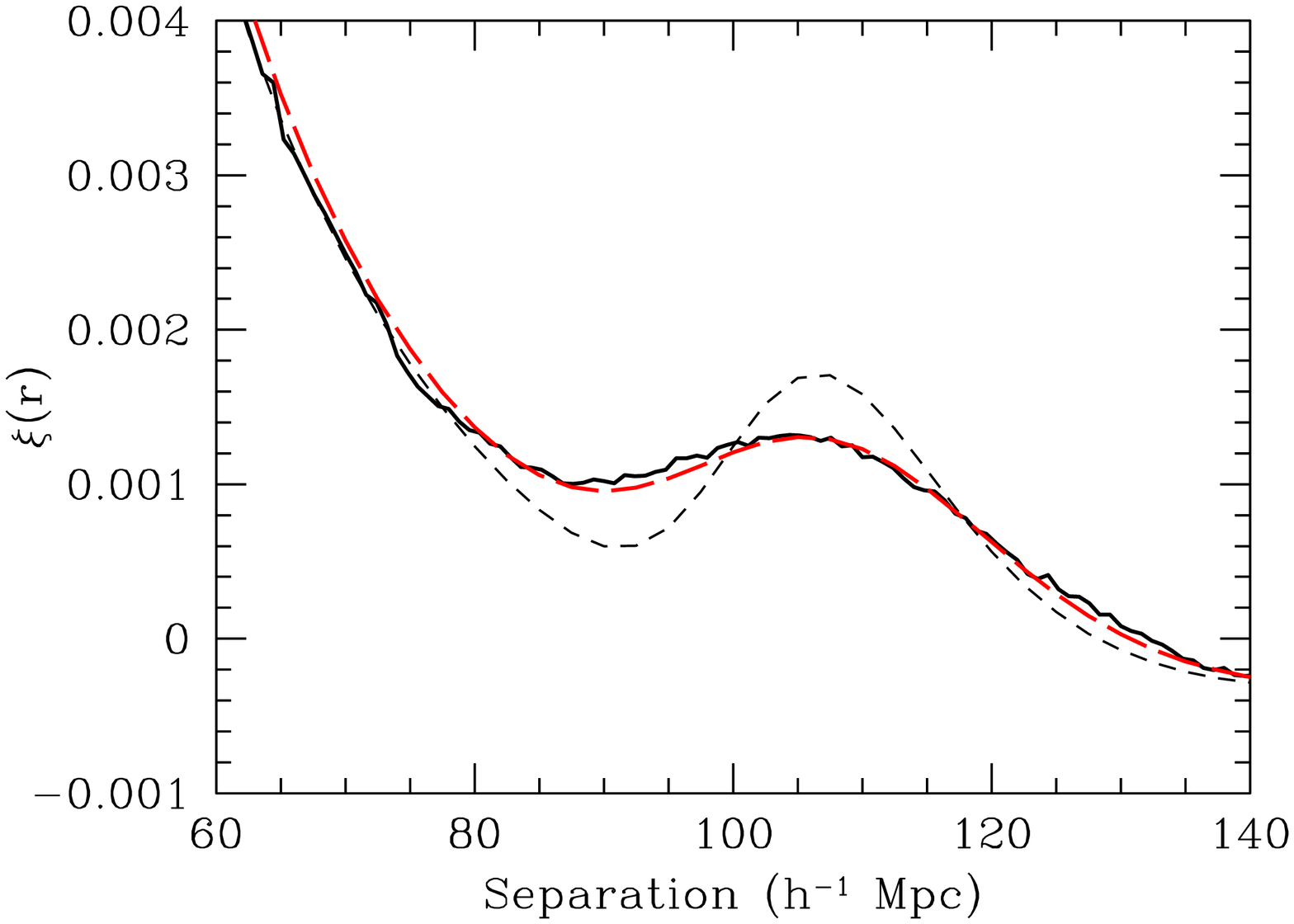}\hspace{\fill}
    \epsfxsize=3in\epsffile{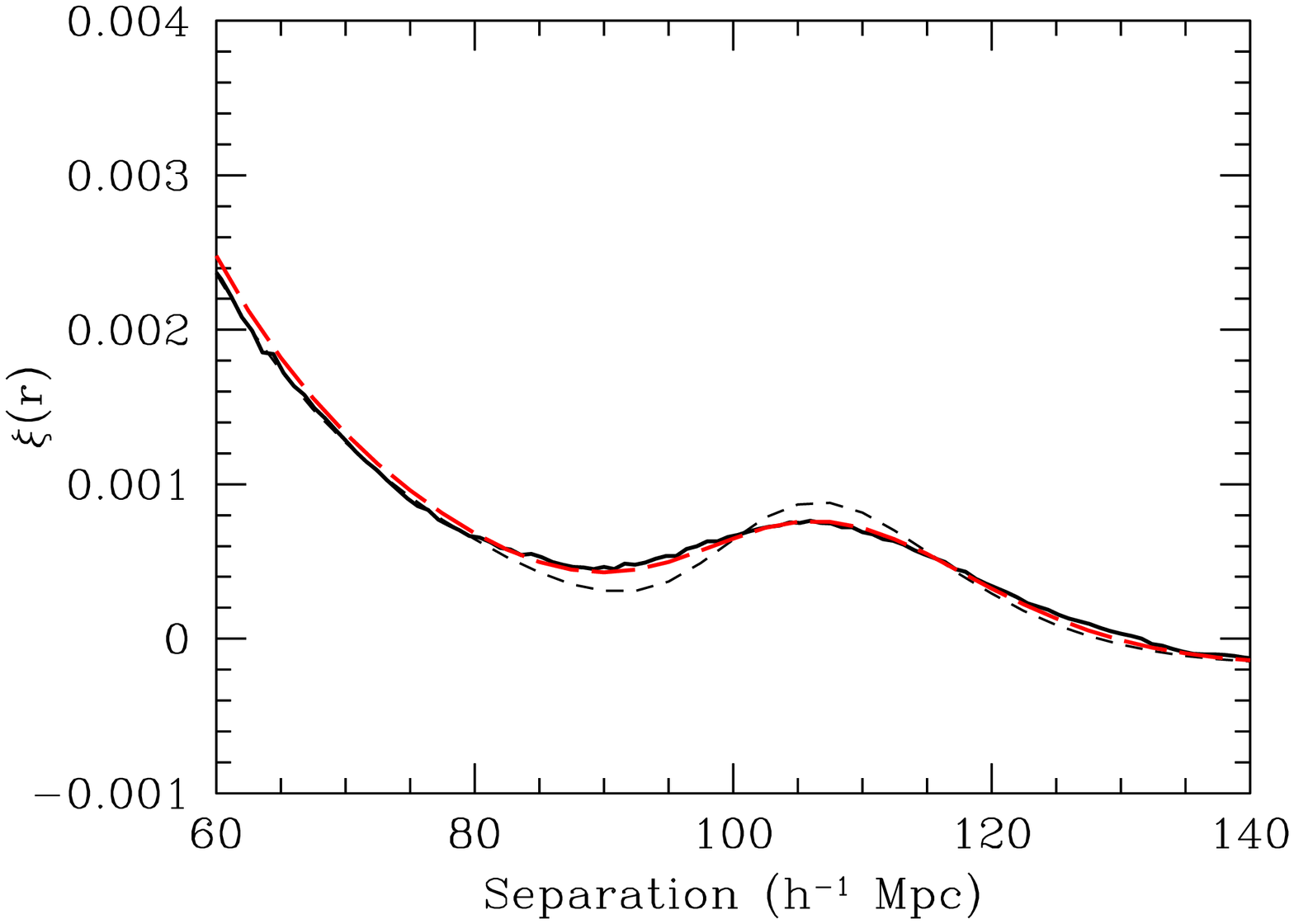}} 
\caption{\label{fig:xi}%
The correlation function in the simulations compared to the model 
with the small-scale linear power restored.
The left panel shows $z=0.3$; the right, $z=1$.
In both panels, the simulation data is the solid black line,
the linear correlation function is the short-dashed thin line,
and the model correlation function is the long-dashed red line.
We have not removed any broad-band nuisance spectra in making
this figure.
}\end{figure}

Figure \ref{fig:power} shows the comparison of the measured power
in the simulations \citep{Seo05} to the smeared model with the small-scale power
restored.  To make this comparison, we need to subtract off the
smooth non-linear power contribution and divide out a linear bias.
We do this by removing a polynomial with constant, $k^2$, $k^4$, and $k^6$
terms and dividing by a constant, fitting over the range $0<k<0.4\ihmpc$
\citep{Seo05}.
For redshift space, we adopt a smoothing function that is the 
spherical average of the elliptical Gaussian in Fourier space.
In detail, we have in Figure \ref{fig:power} shown the power
divided by the $z=49$ input power spectrum.  Purely linear 
evolution would be a line at unity; degradations of the 
acoustic signature puts oscillations in the plot.

Overall the model is a very good fit.  The amplitude of the 
oscillations in the power spectrum is well matched.  The higher harmonics show more
residuals, and it is not clear how much of this is a failing of 
the model versus noise in the power spectra.  Many of the
excursions at high wavenumber cannot be matched even if the
acoustic oscillations are completely erased; i.e., they must
be non-Gaussian noise in the simulated power spectra.
However, the model assumptions to ignore the small deviations
from a Gaussian distribution of displacements, the 
transverse displacements, and the scale dependence of the rms
displacement could result in minor changes to the residuals.
It is important to note that even $3.75\hgpcC$ still leaves 
enough sample variance in the power spectrum that one cannot
fully test this model at the level of 1\% residuals.

Figure \ref{fig:xi} shows the match between the 
correlation function in the simulation and that of the model
with the small-scale power restored.
Here, we do not remove any smooth power spectrum; all of these
nuisance terms appear at small separations in the correlation 
function.
Again, the agreement is good: the model seems to correctly 
match the smearing of the peak in the correlation function.

\section{Numerical Results for Biased Tracers}  \label{sec:biased}

\begin{table*}[t]
\footnotesize
\caption{\label{tab:bias}}
\begin{center}
{\sc Lagrangian Displacements Distribution for Biased Tracers\\}
\begin{tabular}{rccccccccc} 
\doubleline
&   &    &     & \multicolumn{4}{c}{Radial} & \multicolumn{2}{c}{Transverse} \\
&$M$&$z$ & $R$ & Mean & $\sigma$ & Skewness & Kurtosis & $\sigma$ & Kurtosis \\
\singleline
Real-space
&10 & 0.3 &100 & --0.39  & 8.78     & --0.028   & 0.034     & 7.80     & 0.007  \\
&30 & 0.3 &100 & --0.48  & 8.97     & --0.026   & 0.024     & 8.02     & --0.005  \\
&10 & 1.0 &100 & --0.34  & 6.55     & --0.017   & 0.012     & 5.91     & --0.028  \\
\singleline
Redshift-space
&10 & 0.3 &100 & --0.65 & 15.00    & --0.051   & 0.22     & 13.51 & 0.26 \\
&30 & 0.3 &100 & --0.81 & 15.45    & --0.049   & 0.22     & 13.99 & 0.25 \\
&10 & 1.0 &100 & --0.64 & 12.93    & --0.037   & 0.18     & 11.80 & 0.21 \\
\doubleline
\end{tabular}
\end{center}
NOTES.---%
Masses of halos are listed as the number of particles $M$;
each particle has a mass of $6\times 10^{11}h^{-1}\msun$.
Separations $R$ as well as mean and rms displacements are given in 
comoving $\hmpc$.  Skewness and kurtosis statistics are the usual
dimensionless normalization.  Negative skewness means that the heavier
tail is inwards. 
For redshift space, we list the displacements in the line-of-sight
direction when this direction is radial to the initial
separation vector.  The displacements in the direction perpendicular
to the line-of-sight are the same as in real space.
\end{table*}

We next turn to biased tracers.  We use a very simple model of bias,
described in \citet{Seo05}.  We find halos with a friends-of-friends
algorithm \citep{Dav85}, place a threshold on halo multiplicity, and 
include all of the particles in the halo as valid tracers.  This is
equivalent to a halo occupation model in which the number of galaxies
in a halo is proportional to the halo mass, if above some threshold,
and all galaxies trace the velocity dispersion of the dark matter.
This is a relatively extreme model and is not intended to be realistic
but simply to explore the basic effect.

\begin{figure*}[t]
\centerline{\epsfxsize=3in\epsffile{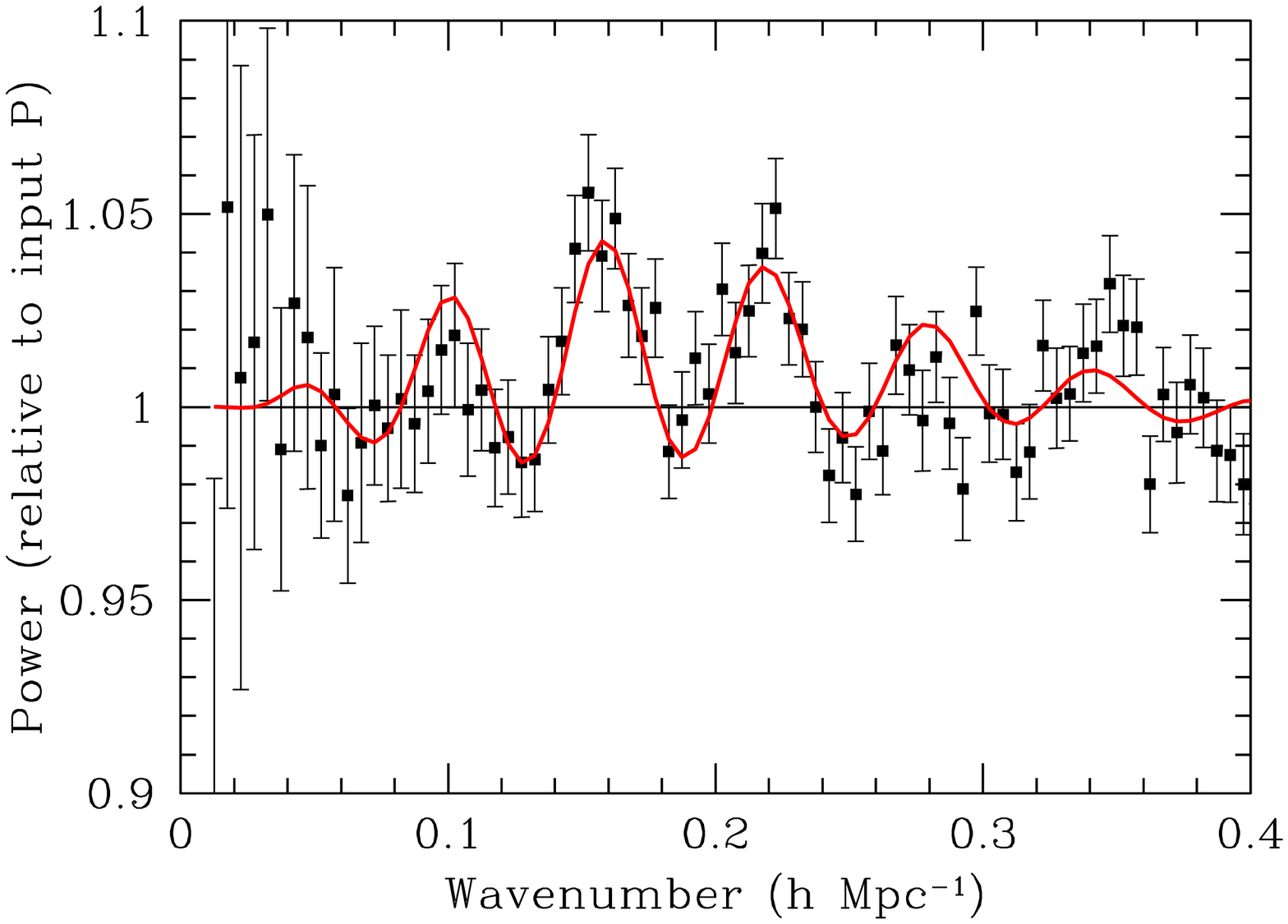}\hspace{\fill}
    \epsfxsize=3in\epsffile{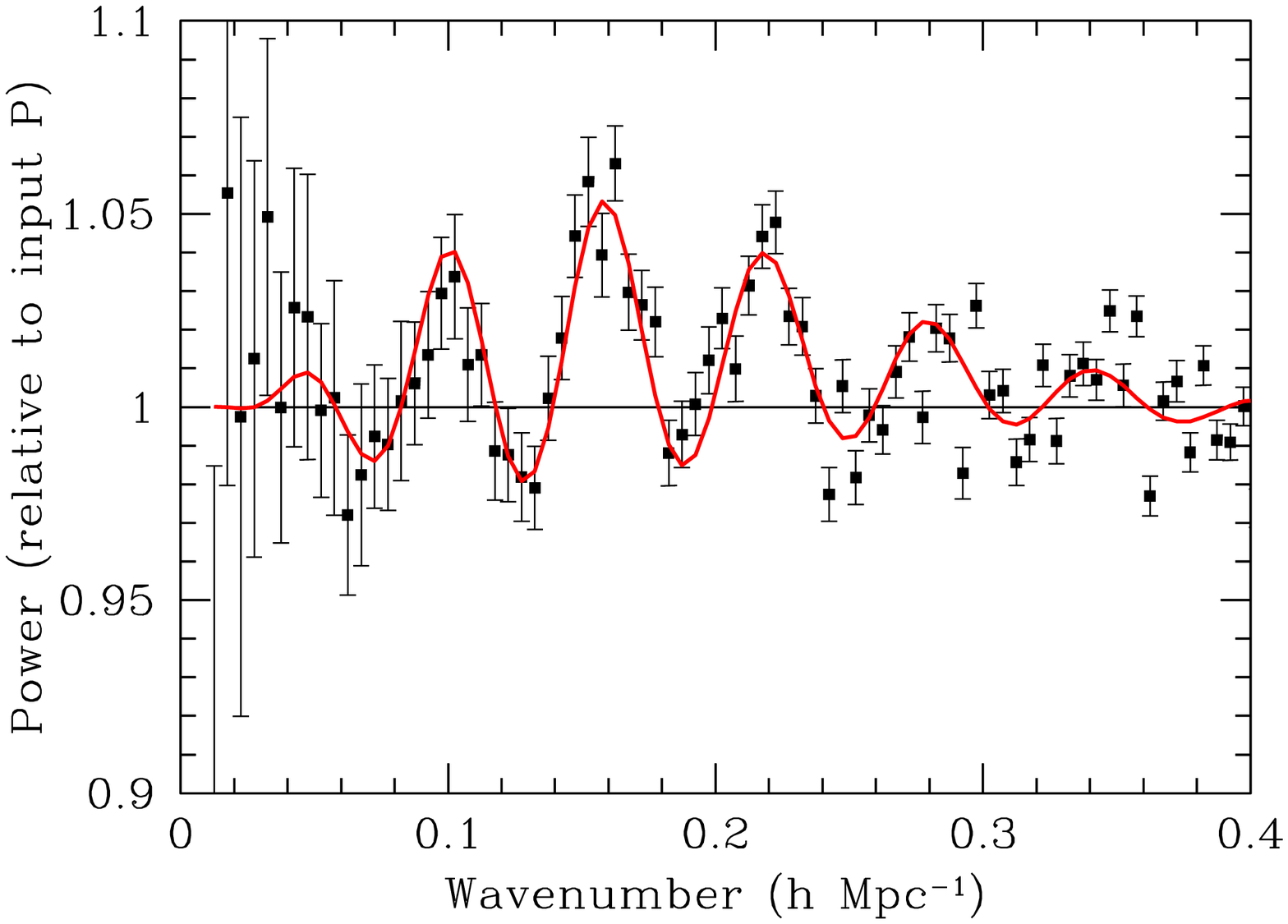}} 
\caption{\label{fig:powerbias}%
As Figure \protect\ref{fig:power}, but for a biased tracer ($M=10$) at $z=0.3$.
The left panel is in real space; the right panel is in redshift space.
The biased power spectrum is noisy compared to the matter power spectrum
because of additional shot noise.  However, one sees that the model
remains a good match to the data.
}\end{figure*}

Table \ref{tab:bias} lists the moments of the Lagrangian displacement
distributions for biased tracers of different mass threshold and 
redshift.  These mass thresholds at $z=0.3$ give large-scale bias
similar to that of the SDSS Luminous Red Galaxy sample \citep{Eis01,Zeh05}.
The primary conclusion is that the bias doesn't change the 
variance of the distribution much.  The rms increases by 10\%
in real space and somewhat more in redshift space.  The latter is
not surprising, because the high halo masses have more small-scale
thermal velocities, particularly in the bias model we use here.  
We interpret the small increase in real space as
indicating that bulk flows affect all galaxies more or less equally,
with the subdominant trend that the high mass halos have more action 
in their small-scale environments.  It appears that bias has rather
little effect on the non-linear degradation of the acoustic peaks.
Figure \ref{fig:powerbias} shows the power spectrum from the $z=0.3$,
$M=10$ biased tracer, along with the model spectrum.  As before, the
agreement is excellent within the errors.

However, we do find in the biased case that the mean pairwise
displacements are no longer zero, as they were for the matter.  The
mean displacements are small, about $0.4\hmpc$ in real space for
separations of $100\hmpc$, somewhat larger in redshift space.  This
would correspond to a 0.4\% bias in the recovered distance scale, 
although it is not clear that this mean displacement is in fact the
correct statistic to predict the exact shift.
The mean for the matter displacements was zero because of homogeneity,
but setting the tracer density according to the late-time matter
density breaks the homogeneity assumption.  We will explore this
effect in detail in the next section.


\section{Non-linear shifts of the acoustic scale}
\label{sec:shift}
\subsection{Spherical collapse}

Until this point, we have been mostly studying how the acoustic
signature is washed out by non-linear structure.  This addresses
the precision with which the acoustic scale can be measured in a 
given survey.  We now turn to the question of how much non-linear collapse can
actually shift the scale of the peak, so as to create a systematic
bias in the measurement.  In detail, the question of a shift 
cannot be separated from the question of what statistic or analysis
one plans to use to measure the acoustic scale.  That is, measuring
the maximum of the peak or some centroid or performing a likelihood
analysis with different templates and nuisance parameters could 
yield different answers when affected by nonlinearities.  The
choice of a measurement scheme is beyond the scope of this paper.
Here we offer a discussion of the mean flows on 150 Mpc scale,
which should explore the dynamics of the non-linearities.

We again begin by considering the configuration-space
picture.  A peak at one location creates a faint spherical echo
and a preferred separation of galaxies.  It is clear that the 
nonlinear motion caused by clusters and flows will be largely independent
between the two galaxies in a given pair.  In other words, most
of the non-linear blurring of the peak (or erasure of the high
harmonics in the power spectrum) does not shift the pairs systematically
to a different separation.  But it is clear that peaks in the 
Universe separated by $\sim\!150$~Mpc will tend to fall towards 
each other, and that voids will
tend to move away from each other.  Hence, there is a small anisotropy
that could shift the acoustic scale.

How big is this effect?  Let us first consider spherical collapse.
If there is a matter overdensity at one point, there will tend to be
a matter overdensity inside a co-centered sphere of radius $R=150\mpc$.  The mean value
of this overdensity is related to the $J_3$ integral of the correlation 
function:
\beq
  \left\langle\delta(<R)\right\rangle = {3\delta_0\over R^3 \sigma_0^2}
  \int_0^R dr\;r^2 \xi(r)
  \equiv {3\delta_0\over \sigma_0^2} J_3(R)
\eeq
Here, $\delta_0$ is the overdensity at the center, $\sigma_0^2$ is
the variance in that overdensity, and $J_3(R)$ is the integral of
the correlation function \citep{Pee80}
\begin{equation}\label{eq:J3}
J_3(s) = \int_0^s {r^2dr\over s^3} \xi(r) = 
\int {k^2 dk\over 2\pi^2} P(k) {j_1(ks)\over ks},
\end{equation}
where $j_1$ is a spherical Bessel function.
One can consider the density field to be smoothed, provided that the smoothing
length is much smaller than $R$.  Since $\delta_0/\sigma_0$ will be of order
unity the mean $\delta(<R)$ will scale as $\sigma_0^{-1}$.
If unsmoothed, then $\sigma_0$ is very large and the mean $\delta(<R)$ 
is near zero; this is simply a statement that behavior at a single point
doesn't bias an entire filled sphere very much.  However, if one smoothes
the density field, $\sigma_0$ becomes smaller.  The natural smoothing
scale for this problem is something similar to the width of the acoustic
shell, since that is the size of the central region that adds coherently
to the acoustic effect.  With this smoothing, $\sigma_0$ is about unity
at $z=0$.

The value of $3J_3(150\mpc)$ in the concordance cosmology \citep{Spe06}
at $z=0$ is about $0.007$,
and so the mean $\delta(<R)$ is of this order.
It is worth remembering that this overdensity is dominated
by the correlations in the initial density field carried by the cold
dark matter, not the acoustic effects.  Translating the overdensity into
a change in the radius of the sphere yields $\delta(<R)/3$, so
the scale is changed only by $O(0.003)$.
In the line-of-sight direction, redshift distortions double the 
apparent infall \citep{Pee80,Kai87} in Einstein-de Sitter, less so in low-density
universes.  It seems difficult to push this to be more than a
1\% effect.

The rms overdensity inside a sphere of $150\mpc$ radius is about 0.07
in the standard cosmology at $z=0$.  However, this overestimates the
infall because the small overdensities that dominate the total in the
sphere can be anywhere inside the sphere, not at the center, so that the
infall isn't radial and doesn't affect the entire spherical acoustic echo.

Thus far, we have only discussed the infall around overdensities.
However, there is a equivalent outflow from underdensities.  For the
matter, these cancel to leading order.  
The next order terms do not cancel, and 
we therefore expect the changes in scale to be $O(10^{-4})$.


\subsection{Biasing the Acoustic Scale}

We next study whether applying a local galaxy bias can create a
first-order motion of the acoustic scale.  We wish to compute 
whether two galaxies separated initially by a given large distance
will tend to fall toward each other.  In the Zel'dovich approximation,
the pair-wise displacement of two initial positions, projected along
the initial separation vector, is (from eq.~[\ref{eq:zel}])
\begin{equation}
  u_{12,\parallel} = {\bfu_{12}\cdot \bfr_{12}\over r_{12}} = 
  	\int {d\bfk\over (2\pi)^3} \delta_{\bfk} {\bfk\cdot\bfr_{12}\over ik^2 r_{12}}
  	     \left[ e^{i\bfk\cdot\bfr_1} - e^{i\bfk\cdot\bfr_2} \right].
\end{equation}
It is clear that $\left<u_{12,\parallel}\right> = 0$; one is performing
an average weighted by the initial volume, not final density, and this
cancellation is required by homogeneity.
If we consider $\left<u_{12,\parallel}\delta_1\right>$, then 
we find that this equal to $-r_{12} J_3(r_{12})$, where 
$J_3$ is defined in Equation (\ref{eq:J3}).
Hence, we do find that objects fall towards overdensities and
away from underdensities.  We note that although this is the same
result as from spherical collapse, we did not assume spherical symmetry
here.

We now wish to consider the mean displacement weighted by pairs, i.e.
\begin{equation}
\upair \equiv { \left< u_{12,\parallel} \delta_1 \delta_2 \right>
	\over \left< \delta_1 \delta_2 \right> }.
\end{equation}
We consider these densities to be those of the initial linear density field.
The denominator $\left< \delta_1 \delta_2 \right>$ is simply $\xi(r_{12})$,
which can be zero, but we will work in the limit where $\xi$ is small but
non-zero; we find that the ratio $\upair$ approaches
a finite number as $\xi\rightarrow0$.

In the Zel'dovich approximation, $u_{12,\parallel}$ is linear in the
density field $\delta$ and hence $\upair=0$ for a Gaussian initial
density field, as the numerator is a 3-point function.  The displacement
field will have corrections at orders above first, but this will produce
$\upair$ at order $O[J_3^2(r_{12})]$ or worse.  In other words, there
is a first-order cancellation in the pair-weighted mean displacement,
when one uses an unbiased tracer.

We next consider weighting by the galaxy overdensity field $\delta_g$.  
We assume a local bias model in which $\delta_g$ is a function simply
of the initial matter density $\delta$.  We write $\delta_{g,1} = \delta_g(\delta_1)$.  Now we have
\begin{equation}\label{eq:pairweight}
\upair \equiv { \left< u_{12,\parallel} \delta_{g,1} \delta_{g,2} \right>
	\over \left< \delta_{g,1} \delta_{g,2} \right> }.
\end{equation}
Importantly, the displacement $u_{12,\parallel}$ still depends on the
matter density field, not the galaxy density field.
In the limit that $|\xi(r_{12})|$ is much smaller than the variance at
a single point, $\sigma^2 = \left<\delta_1^2\right>$, the mean Zel'dovich
displacement given $\delta_1$ and $\delta_2$ is 
\begin{equation}
\left< u_{12,\parallel}\right>_{\delta_1,\delta_2}
= - {r_{12} J_3(r_{12}) \over \sigma^2} \left(\delta_1 + \delta_2\right).
\end{equation}
Keeping only the leading-order terms, we have
\begin{equation}
\upair \equiv - {r_{12} J_3(r_{12}) \over \sigma^2} 
    { \left< (\delta_1+\delta_2) \delta_{g,1} \delta_{g,2} \right>
	\over \left< \delta_{g,1} \delta_{g,2} \right> }.
\end{equation}

We next consider the form of the function $\delta_g$.  We must have
$\left<\delta_g\right>=0$, where the averaging is over the distribution
of the matter density.  This simply defines the mean density of galaxies.
We also have $\left<\delta_g \delta \right> = b\sigma^2$,
where $\sigma^2 = \left<\delta^2\right>$ and $b$ is the familiar large-scale
bias.  Allowing $\delta_g$ to be a stochastic distribution depending on
the matter density at that point \citep[e.g.][]{Dek99}
doesn't alter the result; to leading 
order, only the mean galaxy density as a function of $\delta$ enters.

To compute the expectation values, we must include the fact that 
$\delta_1$ and $\delta_2$ are correlated.  In the limit that 
$|\xi(r_{12})|\ll \sigma^2$, we find
$\left< \delta_{g,1} \delta_{g,2} \right> = b^2\xi(r_{12})$
and 
$\left< \delta_1 \delta_{g,1} \delta_{g,2} \right> = 
\left[b\xi(r_{12})/\sigma^2\right] 
\left< \delta_{g,1} \delta_1^2 \right>$.
If we define rescaled density fields $\nu = \delta/\sigma$ and
$\nu_g = \delta_g/b\sigma$, then we reach the result
\begin{equation}
{\upair\over r_{12}} = -{2 J_3(r_{12})\over \sigma} \left< \nu^2 \nu_g(\nu)\right>
\end{equation}
where the expectation value is computed with $\nu$ being distributed as a 
Gaussian of unit variance.  The function $\nu_g$ is also subject to 
$\left<\nu_g\right>=0$ and $\left<\nu_g\nu\right>=1$, the latter being
due to the scaling by $b$.

Hence, the fractional change in the pair-weighted separation can be
first order in the large-scale correlation function if the local
bias is suitably chosen.
However, for a linear bias, $\nu_g\propto\nu$ (which must actually be
$\nu_g=\nu$), the first-order effect cancels.  We remind the reader
that $2 J_3(r_{12})$ is about 0.5\% for the standard cosmology at
$z=0$ (and of course smaller at high redshift).

The pair-weighted mean displacement of Equation \ref{eq:pairweight} 
is not the same as the mean displacement between pairs of tracers,
as studied in \S~\ref{sec:biased}.  The difference is effectively that
of weighting by $\delta_g$ or by $(1+\delta_g)/(1+\delta_m)$.  
These clearly disagree
on the relative weighting of overdense and underdense regions, and the
tracer-based counting may fail to include the voids properly.
However, in the 
limit of a bias model in which galaxies form only above a high density
threshold, the two calculations converge to the same answer.

For a toy model inspired by Press-Schechter (1974) theory, we 
consider a model in which galaxy formation obeys a simple density
threshold $\delta_t$, typically $\sim\!1.7$.  
For a threshold $\nu_t\equiv \delta_t/\sigma$,
we find $\left< \nu^2 \nu_g\right> = \nu_t$.  This yields 
${\upair/r_{12}} = -J_3(r_{12}) (2\nu_t^2/\delta_t)$.  
The usual problems in accounting for underdense regions in Press-Schechter
theory recommend that one only consider this model for reasonably
extreme peaks, i.e. $\nu_t\gtrsim2$.
For a fixed
$\delta_t$ and smoothing scale (and hence a rapidly declining number density
of objects at higher redshift), the redshift dependence of $\nu_t^2$ and
$J_3$ cancel, matching the behavior in Table \ref{tab:bias}.
Matching the bias models used at $z=0.3$ in \S~\ref{sec:biased} would
suggest $\nu_t\approx1.6$, which makes $\upair/r_{12}\sim -0.6\%$,
somewhat larger than the measured mean displacement but in
qualitative agreement.  However, it is not surprising that the actual halo
bias is closer to linear than the sharp threshold model.

Estimating an exact level of the bias in the measured acoustic scale
is not straight-forward from this local bias formalism, as one has to decide what smoothing scale to
apply to the initial density field, both to define the local bias model
$\nu_g(\nu)$ and to compute the matter variance $\sigma^2$.  One sees,
however, that one is caught between two limits in which the cancellation
is restored: if one chooses a large smoothing scale, then $\sigma$ can 
be small, but the local bias becomes more linear, whereas with a small
smoothing scale, the bias will be non-linear, but $\sigma$ is larger.
Hence, it seems likely that the combination $\left< \nu^2\nu_g\right>
\sigma^{-1}$ will be somewhat less than 1 at $z=0$, yielding a shift of
about 0.5\%.  The shift in the acoustic scale should decrease at high
redshift, somewhat slower than the square of the growth function, for tracers
of a given number density.

To summarize, for objects that trace the matter, the inflow between
positive density pairs exactly cancels the outflow between negative
density pairs.  Galaxy bias can shift the balance, yielding a net flow.
Fortunately, the effect involved, namely the flows on 150 Mpc scale 
is very easy to simulate.  We expect that large volume gravitational
simulations with relatively poor mass resolution should be able to give
accurate estimates of the effect for various galaxy bias prescriptions.

\section{Conclusions}  \label{sec:conclusions}

We have investigated the non-linear degradation of the baryon
acoustic signature through several different methods.  We argue
that as the clustering signature is manifested as a peak in the
correlation function on large scales, the degradation enters primarily
through the motion of the two galaxies altering the separation
between them.  This smearing is due to small-scale thermal
motion, e.g., cluster formation, and coherent
motions, e.g., bulk flows into superclusters or out of voids between
the two objects.  This smearing broadens the peak in the correlation
function and decreases the higher harmonics in the power spectrum.

We model this smearing by measuring the distribution of differences 
of the Lagrangian displacements of pairs of particles initially
separated by a separation equal to the sound horizon.
These distributions are close to Gaussian in both real and redshift space,
and we find that the redshift dependence of the rms width is well 
predicted by the linear-theory scaling predictions.
We propose convolving the linear theory correlation function with this
Gaussian.  The linear theory power spectrum can be multiplied
by the Fourier transform of this Gaussian, and the small-scale
power restored using the no-wiggle form from \citet{Eis98}.

This model has significant advantages.  First, the displacements
can be calculated with a relatively modest volume of cosmological
simulations compared with those required to measure the degraded
power spectrum.  However, one should use box sizes of at least
$512\hmpc$ to model the required wavelengths.  Second, the Lagrangian
displacements are easy to calculate in real or redshift space and
can be applied to arbitrary tracers of the field, as the fuzzy
nature of the initial position of a galaxy is still well confined
relative to the width of the Lagrangian displacement distribution.
Third, in the context of survey parameter estimation, one can now
avoid the assumption of simply truncating the linear mode counting
at a given maximum wavenumber, as has been the standard practice
\citep[e.g.,][among many others]{EHT98,Seo03}.  Instead, we now have an accurate
model for the quasi-linear acoustic signature, in which the higher
harmonics are gradually erased.  We plan to include this model in
survey forecasts in a future paper.

Although degradations of the acoustic signature are important to
estimating the statistical performance of a given volume, an actual
shift in the acoustic scale would be more important, as this would
lead to a bias in the inferred distance scale.  The idea that the
acoustic signature is a single peak in the correlation function
allows one to argue that these effects must be small: to shift the
peak, one must systematically move pairs initially at $150\mpc$ scale
outward or inward on average.  We have investigated this with two
different models, the mean pairwise displacement of tracers and
the pair-weighted mean displacement in the Zel'dovich approximation.
In both cases, we find that linearly biased tracers imply a 
cancellation of the first-order terms, which would make the shift
negligible at $O(10^{-4})$.  With biased galaxy formation, we estimate
shifts of order 0.5\% for objects with $b\approx2$ at $z=0$, dropping at higher redshift.
However, neither of our models are exact, nor have we considered the
issue of how a particular choice of statistic by which one measures
the acoustic scale might be affected by non-linearities.
We believe that 
large volumes of simulations will be needed to refine the estimates
and calibrate particular measurement techniques.
It is highly plausible that corrections could be derived from simulations
that would decrease the residual shift by a factor of a few
and reach $10^{-3}$ accuracy.  After all, any survey capable of 
reach statistical precision on the acoustic scale at levels of 0.5\%
or better will have fantastic data on the small-scale clustering
and environments of the tracer galaxies.

By viewing the non-linear acoustic signature as the spreading of a preferred 
separation, one sees that the acoustic scale must be robust to the
effects of scale-dependent bias and non-linear clustering.
This is not obvious in the power spectrum, where the higher
acoustic harmonics ($k\approx0.2\ihmpc$) appear at the same
wavenumber as the quasi-linear regime.  The key point is that
the peaks and troughs of the acoustic series define a beat 
frequency that is at very large wavelength.  Alternatively 
stated, non-linearities on $10\hmpc$ scales can only introduce
broad-band ($\Delta k\approx 0.1\ihmpc$) modifications of the power
spectrum and therefore cannot affect the peaks and troughs differentially,
save to damp out the entire linear spectrum.

The fact that small-scale non-linear clustering produces smooth power
spectra compared to the acoustic oscillations is also important for
making unbiased measurements of the acoustic scale.  The measured power
spectra will generally be tilted relative to the linear spectrum, and
it is well-known that when measuring the position of a peak, one must
avoid tilting the baseline on which the peak sits.  This leads to
the speculation that non-linearities can bias the measurement of the
acoustic scale.  We see here that this need not be the case.  In 
configuration space, these effects collapse to small separations,
well-separated from the acoustic scale.  Alternatively, in Fourier
space, one can marginalize over smooth nuisance functions.  The 
linear power spectrum predicted as part of the sound horizon calculation
provides an easy template against which to make an unbiased measurement.

In short, the acoustic scale is a robust standard ruler because it is
fundamentally a $150\mpc$ clustering signature.  This is far larger
than the characteristic scales for halo formation and galaxy bias,
and the 10\% width of the peak makes it very implausible to mimic by
astrophysical processes.  We have shown here that the degradations of
the signature have benign quasi-linear causes.  These can be computed
to high accuracy with N-body gravitational simulations.  The remaining
challenge for the method is the enormous survey volumes required to measure
the scale to high precision.

\bigskip

DJE thanks Michael Joyce for useful conversations
and the Miller Institute at the University of California
for support during a key period in the development of this paper.
DJE and HS are supported by grant AST-0407200 from the 
National Science Foundation.  
DJE is further supported by an Alfred P. Sloan Research Fellowship.
M.W. is supported by NASA and the NSF.


\appendix

\section{An Investigation of Local Bias}  \label{sec:bias}

In the configuration-space picture of the acoustic signature, the sound waves
created by a peak in the density field leave a small residual overdensity
in a shell at $150\mpc$ radius.  
We consider in this appendix whether there is a judicious choice
of tracers that would increase the contrast of the shell or peak,
so as to increase the strength of the signature in the correlation
function.  Unfortunately, we will conclude that there is not, 
reinforcing the concepts of linear bias.

The thickness of this shell is 
about $10\mpc$ because of Silk damping and the damping of oscillatory
modes.  Hence, two points separated by $\ll 10\mpc$ create acoustic
shells that essentially overlap.  In what follows, we consider the
density field smoothed on $10\mpc$ scales.

The variance at low redshift in the overdensity in $10\mpc$ patches 
is roughly unity.  The corresponding echos at $150\mpc$ separation are
only 1\% in amplitude, far smaller than the primary variations at those
locations (which are again order unity).  The smallness of the echo
suggests that a local bias model should apply.  We therefore consider
the response of the behavior in $10\mpc$ patches when acted upon by
bulk shifts in their density, corresponding to long-wavelength 
perturbations.

We will consider two regions of size $\sim\!10\mpc$ separated by a 
distance of $150\mpc$.  
We write the linear matter overdensity (ignoring the acoustic
effect) in the two as $\delta_{m,1}$ and $\delta_{m,2}$; 
The $\delta_m$ will be Gaussian distributed with variance $\sigma$;
we write the distribution as $p(\delta_m)$.  
Because of the acoustic effect, the density at location 2 will
actually be $\delta_{m,2}+\alpha\delta_{m,1}$, where $\alpha$
is about 0.01.  There is a recipricol effect on the density at
location 1 which we ignore for clarity.

Now we imagine that galaxies are formed in a region of density 
$\delta_m$ with a mean density of $\rho_g(\delta_m)$.  Of course,
the actual number will scatter from the mean, but only the mean
enters this calculation.  The homogeneous density of galaxies
is $\bar\rho_g = \int d\delta_m\;p(\delta_m)\rho_g(\delta_m)$.
We define the galaxy overdensity as 
$\delta_g(\delta_m) = [\rho_g(\delta_m)-\bar\rho]/\bar\rho$.

With such a model, the bias at large scales (small wavenumber)
should be the response of this density to a small long-wavelength
fluctuation in the matter density:
\begin{equation}\label{eq:bias}
b = \lim_{\epsilon\rightarrow0} 
{\int d\delta_m \;p(\delta_m)\rho_g(\delta_m+\epsilon) - \bar\rho
\over \epsilon \bar\rho}
= {\int d\delta_m\;p(\delta_m) \delta_g(\delta_m+\epsilon)\over\epsilon}
= \int d\delta_m\;p(\delta_m) {d\delta_g\over d\delta_m}.
\end{equation}
Meanwhile, the amplitude of the acoustic effect should be the
product of the galaxy overdensities at the two locations:
\begin{equation}
B^2 = \int d\delta_{m,1}\;p(\delta_{m,1}) \delta_g(\delta_{m,1})
\int d\delta_{m,2}\;p(\delta_{m,2}) \delta_g(\delta_{m,2}+\alpha\delta_{m,1})
\end{equation}
Because $\alpha$ is small, we can use the limit in equation (\ref{eq:bias})
to form
\begin{equation}\label{eq:B}
B^2 = \int d\delta_{m,1}\;p(\delta_{m,1}) \delta_g(\delta_{m,1}) 
b \alpha \delta_{m,1}.
\end{equation}

One might at this point have thought that by picking $\delta_g$ judiciously
one could produce an amplitude for the acoustic peak that would 
increase its contrast relative to the square of the large-scale bias.
In other words, one has a large peak in region 1 and a faint echo in 
region 2.  The galaxy density in region 2 is unavoidably controlled
by the standard large-scale bias, as the acoustic signature is a 
subdominant echo at that location.  But one could hope that one could pick
a tracer that would emphasize region 1 more.  One could even imagine
cross-correlating two tracer sets, one that maximizes signal-to-noise
for small long-wavelength perturbations (region 2) and the other that
picks out high peaks (region 1).

Unfortunately, this is all for naught.  Integrating equation (\ref{eq:B})
by parts yields
\beq
B^2 = \alpha b^2 \sigma^2
\eeq
for any $\rho_g(\delta_m)$ bias function.  The trick is that 
the indefinite integral of $\delta_m p(\delta_m)$ is just $-\sigma^2 p(\delta_m)$
because the initial density distribution $p$ is a Gaussian.

This result shows that at the level of local bias, one cannot
pick the tracer to maximize the contrast in the acoustic peak
in the linear regime.  This is a simple corollary of the usual
theorems about local bias \citep{Col93,Sch98}.  It may be
possible to find tracers that minimize the non-linear spreading
of the acoustic peak or that minimize the amount of small-scale
near-white noise; such tracers would improve the signal-to-noise
ratio of the measurement.  But there is no opportunity at the 
simple level of pair counting.

{}

\end{document}